\documentclass[reprint, 
 amsmath,amssymb,
 aps,
floatfix
]{revtex4-1}

\usepackage{graphicx}
\usepackage{dcolumn}
\usepackage{bm}
\usepackage{hyperref}
\usepackage{upgreek}
\usepackage{xcolor}
\usepackage[version=3]{mhchem}

\begin{document}

\title{Reversible Ionic Aggregation Kinetics in Concentrated Electrolytes}

\author{Zachary A. H. Goodwin}
\email{zac.goodwin@materials.ox.ac.uk}
\affiliation{John A. Paulson School of Engineering and Applied Sciences, Harvard University, Cambridge, Massachusetts 02138, United States}
\affiliation{Department of Materials, University of Oxford, Parks Road, Oxford OX1 3PH, United Kingdom}

\date{\today}

\begin{abstract}
Here we develop a formalism for reversible ionic aggregation kinetics in an example concentrated electrolyte\textcolor{black}{, building on previous equilibrium work of McEldrew and co-workers, and thermoreversible polymers and patchy particle systems.} This is achieved through solving a macroscopic rate equation of open/occupied association sites, \textcolor{black}{shown to be} a solution of the reversible Smoluchowski aggregation equation\textcolor{black}{, which predicts how ionic associations in electrolytes change subject to a step-change in conditions}. We \textcolor{black}{test} the derived equations against atomistic molecular dynamics simulations of a salt-in-ionic liquid\textcolor{black}{, where} good qualitative agreement is obtained, but quantitative differences are found. \textcolor{black}{This} highlights the multiple time scales that exist in concentrated electrolytes\textcolor{black}{, with a fast timescale preceding a longer timescale}. We hope this formalism \textcolor{black}{opens new avenues in understanding the dynamics and non-equilibrium behaviour in electrolytes}. \textcolor{black}{For example, the formalism can be developed} further to investigate the non-Newtonian behaviour of concentrated electrolytes, double layer charging, and the slow dynamics of these electrolytes in confinement. 
\end{abstract}

\maketitle

\section{Introduction}

Concentrated electrolytes, such as water-in-salt electrolytes (WiSES)~\cite{Suo2013,Suo2015,Wang2016,Wang2018,Yamada2016,suo2017water,dou2018safe}, ionic liquids (ILs)~\cite{Kornyshev2007,Fedorov2014,Welton1999,Welton2011}, salt-in-ILs (SiILs)~\cite{molinari2019general,molinari2019transport,molinari2020chelation,Zhang2024}, amongst many other examples~\cite{efaw2023localized,xu2014electrolytes,Cheol2023HE,Xie2023}, are increasingly of interest in various \textcolor{black}{technologies}~\cite{xu2004nonaqueous,Goodenough2013ThePerspective,xu2014electrolytes,li2020new,Fedorov2014,Welton1999,Welton2011,Gebbie2023,Zheng2017Uni,phelan2024role,Phelan2025Effect}. For example, ILs have potential applications as `green' solvents for electrochemical reactions at electrified interfaces, and electrolytes for supercapacitors owing to their low vapour pressure and flammability, and because of their wide electrochemical stability window (ESW)~\cite{Kornyshev2007,Fedorov2014,Welton1999,Welton2011}. Similarly, WiSEs are promising alternative electrolytes for batteries which are non-flammable, owing to water being the solvent, but still have a wide ESW \textcolor{black}{from} the high salt concentration~\cite{Suo2015,yu2020asymmetric,borodin2017liquid,Lannelongue2018,vatamanu2017}. Finally, SiILs, or salt-doped ILs, are also finding applications in energy \textcolor{black}{technologies}, where the unique properties of ILs are being exploited as the `solvent' to dissolve active ions~\cite{Chagas2019,lourenco2021theoretical,kuhnel2017}. 

These \textcolor{black}{concentrated} electrolytes also pose interesting fundamental questions~\cite{borodin2020uncharted,Zheng2017Uni,Zhou2022Agg}. For example, in ILs~\cite{Gebbie2013,Gebbie2015,Han2020}, WiSEs~\cite{Han2021WiSE,groves2021surface} and SiILs~\cite{Zhang2024} (\textcolor{black}{and also other concentrated electrolytes}~\cite{perez2017underscreening}), the observation of ``anomalous underscreening'' \textcolor{black}{(unexpectedly extremely long-ranged force decay lengths in surface force balance/apparatus measurements)} raised questions about \textcolor{black}{the electrostatic correlations in bulk, equilibrium concentrated electrolytes}~\cite{Gebbie2013,Gebbie2015,Han2020,smith2016electrostatic,perez2017underscreening,Han2021WiSE,Zhang2024}. While recent research has indicated these are non-equilibrium measurements~\cite{Cross2026}, this motivated \textcolor{black}{significant} work into understanding \textcolor{black}{bulk, equilibrium} concentrated electrolytes~\cite{hartel2023anomalous,Kjellander2016,avni2020charge,adar2019screening,Pedro2020,krucker2021underscreening}. Furthermore, WiSEs behave \textcolor{black}{as} non-Newtonian \textcolor{black}{fluids}~\cite{Yamaguchi2024nn}, and have nano-channels of ionic aggregates interpenetrated by water domains \textcolor{black}{that result in} facile transport of the active cations~\cite{yu2020asymmetric,lim2018,borodin2017liquid,Lannelongue2018,vatamanu2017}. In the context of SiILs, negative transference numbers of alkali metal cations \textcolor{black}{at low mole fractions} generated much interest in studying the transport properties of concentrated electrolytes~\cite{molinari2019general,molinari2019transport}.

These properties have been understood, at least in part, from the ionic aggregates and solvation environments in the \textcolor{black}{concentrated} electrolytes~\cite{borodin2020uncharted,Zheng2017Uni,Zhou2022Agg,Shirley2022}. For example, the negative transference numbers of SiILs can be explained through the formation of a negative shell of anions around the alkali metal cations~\cite{molinari2019general,molinari2019transport}. Moreover, more generally, ionic associations and solvation is known to be important in understanding interphase formation in batteries~\cite{Borodin2014SEI,Borodin2017Mod,Qisheng23JACS} \textcolor{black}{and} the kinetics of electrocatalysis reactions~\cite{Gebbie2023}, \textcolor{black}{ for example}. Typically, these \textcolor{black}{conclusions} have been \textcolor{black}{drawn} from \textcolor{black}{correlating} the coordination environments of the electrolytes \textcolor{black}{(and from the relative numbers of solvent separated ion pairs, contact ion pairs and ``aggregates'')} \textcolor{black}{, and these observed properties.} Such approaches, however, provide little fundamental insight into the underlying laws which link these coordination environments to other properties of the electrolytes.

There are, however, microscopic theories which directly link the ionic associations and solvation to physiochemical properties of these electrolytes~\cite{Tkachenko2025,mceldrew2020theory}. A notable example is the thermoreversible ionic aggregation model developed by McEldrew \textit{et al.}~\cite{McEldrewPhD,mceldrew2020theory,mceldrew2020corr,McEldrewsalt2021,mceldrew2021ion,Goodwin2023}, which links coordination environments to activity \textcolor{black}{and redox potentials}~\cite{Phelan2025Effect}, transference numbers~\cite{mceldrew2020corr,mceldrew2021ion,McEldrewsalt2021}, electrical double layers~\cite{Goodwin2022EDL,Goodwin2022Kornyshev,Markiewitz2025,markiewitz2024} and many more properties~\cite{mceldrew2020theory}. This \textcolor{black}{framework} is founded on the classical polymer theories of Flory, Stockmayer and Tanaka~\cite{flory1942thermodynamics,stockmayer1943theory,stockmayer1944theory,stockmayer1952molecular,ishida1997,tanaka1989,tanaka1990thermodynamic,tanaka1994,tanaka1995,tanaka1998,tanaka1999,tanaka2002}, which describe the thermoreversible \textcolor{black}{bonding} of monomer units into a polymer, where McEldrew \textit{et al.}~\cite{mceldrew2020theory} translated this to the thermoreversible association of ions into aggregates (\textcolor{black}{i.e.,} an alternating copolymer of cations and anions, decorated by solvent). 

While McEldrew \textit{et al.} have mainly investigated equilibrium and linear-response transport properties~\cite{mceldrew2020theory,mceldrew2020corr,mceldrew2021ion,McEldrewsalt2021,Goodwin2023}, in the context of polymers and patchy particle systems~\cite{Sciortino2011CO,Sciortino2009PF,Corezzi2012,groschel2013,Umar2013,Zaccone2013,Zaccone2011}, non-equilibrium properties have \textcolor{black}{also} been investigated. For example, Sciortino, Tartaglia and co-workers connected the kinetics of patchy particle aggregation to the distribution of clusters~\cite{Sciortino2011CO,Sciortino2009PF,Corezzi2012,Corezzi2010,Bianchi2007,Corezzi2009,Sciortino2008,Tartaglia2009}, and \textcolor{black}{Tanaka~\cite{Tanaka2024,Tanaka2023}} \textcolor{black}{used this to understand the viscoelastic} response of polymers with thermoreversible associations. This presents itself as a significant \textcolor{black}{opportunity}, where concentrated electrolytes could greatly benefit from drawing on these fields of research. For example, the dynamics of ionic associations are routinely studied (through residence times) of electrolytes~\cite{molinari2019general,molinari2019transport}, and it is known that the desolvation time scale is important for intercalation rates~\cite{Xu2021desol,Rubinstein2026}. However, more quantitatively linking these to \textcolor{black}{measurable electrolyte properties through} a microscopic theory remains lacking~\cite{Tian2022}.

Here, we take the first steps to apply the non-equilibrium formalisms from polymer and patchy particle systems to concentrated electrolytes. Specifically, we investigate the dynamics of reversible ionic aggregation in bulk SiILs. \textcolor{black}{Analytical expressions for how the coordination numbers of the ions vary subject to a step-change in conditions are derived. A single decay time is analytically obtained in the theory, and found to depend on the rate of forming a single association and the final coordination numbers. To validate these derived equations, we perform molecular dynamics simulations.} Overall, qualitative agreement is found between theory and simulations. The quantitative differences between the theory and simulations reveal a single rate constant \textcolor{black}{is not sufficient to} describe the ionic aggregation over all time scales. The formalism provides insight into the kinetics of reversible ionic aggregation in SiILs, \textcolor{black}{and indicates avenues for} further developments of the theory \textcolor{black}{to more quantitatively link the dynamics of associations to physiochemical properties of concentrated electrolytes.}

\textcolor{black}{This paper is structured as follows. First is the Theory Section. In this section, we initially cover the background and assumptions of the formalism of McEldrew \textit{et al.}~\cite{mceldrew2020theory,mceldrew2020corr,mceldrew2021ion,McEldrewsalt2021,Goodwin2023}. Next we go into a review of the bulk theory, providing a high-level overview initially. The remainder of the Theory Section introduces the non-equilibrium formalism from patchy particle systems, but casts it into the context of concentrated electrolytes, and then we solve for how the coordination numbers change in time from a step-change in conditions. Second, we provide the details of the atomistic MD simulations performed to test the derived equations. Third is the Results Section, which is self-contained and compares the non-equilibrium MD simulation results for how coordination numbers change in time against the derived, analytical equations. Finally, we discuss the comparison between theory and simulation, provide details of how the presented formalism can be compared against experiments, and further extensions of this formalism to transport properties, interfacial properties and rheological properties is outlined.}

\section{Theory}

Here we study salt-in-ionic liquids (SiILs) as an example system to investigate the kinetics of reversible ionic aggregation. In particular, we study alkali metal doped ILs, where the anion is the same in each salt (3 species in total). Previously, McEldrew \textit{et al.} developed an equilibrium theory of the bulk ionic aggregates~\cite{McEldrewsalt2021}, quantified the effective charge of the alkali metal cation, and studied the electrical double layer~\cite{markiewitz2024,Zhang2024}. We recommend that the reader familiarities themselves with these previous works~\cite{McEldrewsalt2021,markiewitz2024,Zhang2024}, also for other electrolytes~\cite{mceldrew2020corr,mceldrew2021ion,Goodwin2023,Phelan2025Effect}, since the formalism has a significant learning curve. Nonetheless, we provide a brief summary of the equilibrium theory before progressing onto the kinetics of aggregate formation. \textcolor{black}{Before progressing onto the bulk and non-equilibrium properties, we first outline the assumptions of the theory and possible limitations.} 

\subsection{Assumptions and Definitions}

\begin{figure*}
    \centering
    \includegraphics[width=0.9\linewidth]{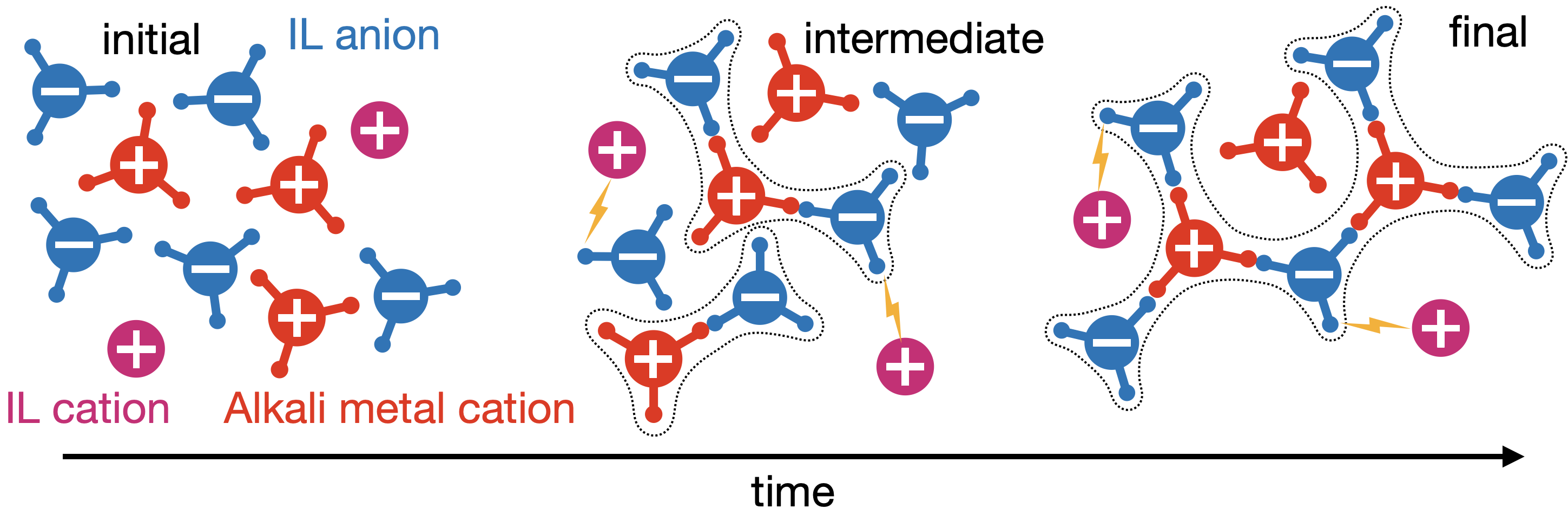}
    \caption{Schematic of non-equilibrium kinetics of ionic aggregation in the studied salt-in-ionic liquid (SIIL). The alkali metal cation and IL anion bind together to form aggregates. The alkali metal cation and IL anions are shown with 3 dangling bonds, to represent they can form associations, while the IL cations are shown without, since they do not associate strongly with the anions. \textcolor{black}{The aggregates formed are denoted by dotted lines surrounding them, while the interactions between the IL cations and anions are shown with a bolt.} From step-changing a property of the system, we investigate how they evolve in time. \textcolor{black}{Initially, the interactions are ``turned off'', but reinstated to then study how the aggregates form.}}
    \label{fig:schem}
\end{figure*}

The alkali metal cations and anions are assumed to form a polydisperse mixture of Cayley-tree aggregates through the associations in their fist coordination shell~\cite{mceldrew2020theory,McEldrewsalt2021}. \textcolor{black}{These aggregates have an alternating cation-anion structure, a reflection of overscreening, as a way of introducing correlations beyond mean-field electrostatics~\cite{avni2020charge,Bazant2011,Goodwin2022EDL}. The Cayley tree approximation of aggregates is central to the analytical tractability of the theory, where long-range order can be constructed from short-range associations~\cite{mceldrew2020theory,McEldrewsalt2021}.} The IL cations are assumed not to participate in ionic associations directly, but can still influence the aggregates through interacting with the \textcolor{black}{open} anion \textcolor{black}{association sites through a regular solution term}~\cite{Welton2025,McEldrewsalt2021,Zhang2024}. This is schematically shown in Fig.~\ref{fig:schem}. \textcolor{black}{All associations and interactions are taken to be independent. We neglect all further interactions beyond these. In the absence of the IL cations, we treat the electrolyte as a `non-interacting' mixture of these polydisperse aggregates.}

We treat the SiIL as an incompressible lattice fluid~\cite{mceldrew2020theory,McEldrewsalt2021}. The alkali metal cations are denoted by $+$, and IL cations denoted by $\oplus$, and anions by $-$. The total volume fractions of ions, $\phi_j$, are known, with $1 = \phi_+ + \phi_- + \phi_\oplus$ from the incompressibility constraint. The lattice site is assumed to be the volume of the alkali metal cation, $v_+$. The volumes occupied by the other ions are expressed relative to this volume through $v_+ \xi_j$, where $\xi_j = v_j/v_+$. The (dimensionless) concentration of each ion per lattice site is given by $c_j = \phi_j/\xi_j$~\cite{mceldrew2020theory,McEldrewsalt2021}. 

The alkali metal cations can form a maximum of $f_{+}$ associations, and the anions a maximum of $f_{-}$ associations. This is referred to as the functionality~\cite{mceldrew2020theory,McEldrewsalt2021}. \textcolor{black}{These association sties, shown as dangling bonds in Fig.~\ref{fig:schem}, represent associations between species. This means that mono-dentate associations and bi-dentate associations, etc., are treated as a single association, since they are between species~\cite{McEldrewsalt2021}.} As these functionalities are larger than 1, a polydisperse cluster distribution can form with clusters of rank $lm$, where $l$ is the number of cations and $m$ is the number of anions, with dimensionless concentration $c_{lm}$~\cite{mceldrew2020theory,McEldrewsalt2021}. For functionalities equal to or larger than 2, a percolating ionic network can emerge~\cite{mceldrew2020theory}. This is referred to as the gel here. In the gel regime, we employ Flory's convention to determine the volume fractions of ions in the sol ($\phi_{+/-}^{sol}$, i.e., not the gel phase) and gel phase ($\phi_{+/-}^{gel}$), where $\phi_{+/-} = \phi_{+/-}^{sol} + \phi_{+/-}^{gel}$. The total dimensionless concentration of alkali metal cations is given  $c_+ = \sum_{lm} l c_{lm} + c_+^{gel}$ and anions $c_- = \sum_{lm} m c_{lm} + c_-^{gel}$. Note $\phi_{lm} = (l + \xi_- m)\tilde{c}_{lm}$ also helps convert between volume fractions and dimensionless concentrations.

\subsection{Equilibrium Theory Overview}

\textcolor{black}{Before getting into the details of the theory, we outline the structure of the derivation and concepts introduced. The assumed free energy functional contains all the information necessary to derive the cluster distribution, $c_{lm}$, which is the central quantity in the theory. From taking functional derivatives of this free energy, and establishing equilibrium between the resulting chemical potentials, we can derive an equilibrium equation for $\phi_{lm}$. From simplifying this, we can arrive at $c_{lm}$. As $c_{lm}$ is a function of the concentration of free species, which we wanted to predict, not be an input, we introduce the concept of association probabilities, and another set of mass action laws for these variables, which allows the full system of equations to be determined, just based on the association constant. }

The free energy functional~\cite{mceldrew2020theory,McEldrewsalt2021,markiewitz2024,Zhang2024} is given by
\begin{align}
    \label{HEnergy} 
    v_+\mathcal{F} =& \int_V  \,d\textbf{r} \dfrac{c_{\oplus}}{\beta} \ln \phi_{\oplus} +\sum_{lm} \left(\dfrac{c_{lm}}{\beta}\ln\phi_{lm} + c_{lm}\Delta_{lm}\right) \nonumber \\
    &+ \int_V  \,d\textbf{r}\chi\phi_{\oplus}\sum_{lm}\left(f_- m -m-l+1\right)c_{lm}  \nonumber \\
    &+ \int_V  \,d\textbf{r}\left( \Delta^{gel}_+c_+^{gel} + \Delta^{gel}_-c_-^{gel} \right),
\end{align}

\noindent where $\beta^{-1} = k_BT$ is inverse thermal energy. The first term is the ideal entropy of the IL cations, and the second term is the ideal entropy from all clusters. The third term is the free energy of forming clusters, where $\Delta_{lm}$ is the free energy of forming each cluster of rank $lm$. The fourth term is the regular solution interaction between the IL cations and open anion binding sites~\cite{McEldrewsalt2021,markiewitz2024,Zhang2024}), with strength $\chi$. The fifth term comes from the free energy of ions associating to the gel, $\Delta_j^{gel}$.

The chemical potential of a rank $lm$ cluster, $\mu_{lm}$, can be determined by \textcolor{black}{the functional derivative of the} free energy with respect to $c_{lm}$. Establishing equilibrium between clusters of rank $lm$ and their free counterparts, we have
\begin{align}
l\mu_{10}+m\mu_{01} = \mu_{lm}.
\label{eq:clusteq}
\end{align}

\noindent Using the \textcolor{black}{expressions for} chemical potential, we obtain the cluster equilibrium
\begin{align}
\phi_{lm}=K_{lm}\phi_{10}^{l}\phi_{01}^{m},
\label{eq:clustp}
\end{align}

\noindent where $K_{lm}=\exp\left\{(l+m-1)(1+\beta\chi \phi_{c_1})-\beta\Delta_{lm}  \right\}$ is the equilibrium constant. \textcolor{black}{To simplify this expression we have to specify $\Delta_{lm}$, which is where our assumed Cayley-tree aggregates enter.} We consider the free energy of forming a cluster of rank $lm$, $\Delta_{lm}$,~\cite{mceldrew2020theory,McEldrewsalt2021} to have three contributions
\begin{equation}
    \Delta_{lm} = \Delta^{comb}_{lm} + \Delta^{conf}_{lm} + \Delta^{bind}_{lm},
\end{equation}

\noindent where $\Delta^{comb}_{lm}$ is the combinatorial entropy \textcolor{black}{(entropy related to the number of ways of arranging those ions in that Cayley-tree aggregate)}, $\Delta^{conf}_{lm}$ is the configurational entropy \textcolor{black}{(related to those aggregates existing on the lattice)} and $\Delta^{bind}_{lm}$ is the binding energy. The combinatorial entropy is given by
\begin{align}
\Delta_{lm}^{comb}=- \dfrac{1}{\beta}\ln f_{+}^lf_{-}^mW_{lm},
\end{align}

\noindent where $W_{lm}$ is \textcolor{black}{related to} the enumeration of the ways that a rank $lm$ cluster can be formed. For Cayley trees, Stockmayer~\cite{stockmayer1943theory,stockmayer1944theory, stockmayer1952molecular} determined the expression to be 
\begin{align}
W_{lm}=\frac{(f_{+}l-l)!(f_{-}m-m)!}{l!m!(f_{+}l-l-m+1)!(f_{-}m-m-l+1)!}.
\end{align}

\noindent The binding energy is approximated as the energy of a single association, $\Delta u$, multiplied by the total number of associations in a cluster \textcolor{black}{(as they are assumed to be independent)}, which for Cayley tree clusters is $l+m-1$. Thus, the binding energy is
\begin{align}
\Delta_{lm}^{bind}= (l+m-1)\Delta u.
\label{eq:bind}
\end{align}

\noindent The configurational entropy describes the entropy of placing a rank $lm$ cluster on the lattice. We modify Flory's expression~\cite{mceldrew2020theory,flory1942thermodynamics,flory1953principles,tanaka1989,tanaka1999} to be
\begin{align}
\beta\Delta_{lm}^{conf} = -\ln\left(\frac{l+\xi_-m}{\xi_-^m}\right) -(l+m-1)(\beta T\Delta s - 1),
\end{align}

\noindent where $\Delta s$ is a per-association entropy change~\cite{Phelan2025Effect}.

From using $\Delta_{lm}$, we obtain the thermodynamically consistent cluster distribution
\begin{align}
c_{lm}=\frac{W_{lm}}{\lambda}  \left(\lambda f_{+}\phi_{10}\right)^l \left(\lambda f_- \phi_{01}/\xi_-\right)^m,
\label{eq:clust}
\end{align}

\noindent where $\lambda$ is the association constant given by
\begin{align}
\lambda=\exp \left\{\beta\left(-\Delta u + T\Delta s + \chi \phi_{\oplus}\right)\right\}.
\label{eq:L}
\end{align}

\noindent \textcolor{black}{The cluster distribution $c_{lm}$ has direct utility, unlike Eq.~\eqref{eq:clustp}, since it essentially only depends one parameter, the association constant, $\lambda$, which needs to be determined.}

In Eq.~\eqref{eq:clust}, $c_{lm}$ is written in terms of the volume fraction of free alkali metal cations ($\phi_{10}$) and anions ($\phi_{01}$)\textcolor{black}{, which also appear to be unknown parameters}. These quantities are, however, not \textit{a priori} known, and moreover, we aimed to predict them from the theory, not have them as inputs for it. To overcome this, we~\cite{ishida1997,tanaka1989,tanaka1990thermodynamic,tanaka1994,tanaka1995,tanaka1998,tanaka1999,tanaka2002,mceldrew2020theory,McEldrewsalt2021} express $\phi_{10/01}$ in terms of $\phi_{+/-}$ and introduce ion association probabilities, $p_{ij}$, which is the probability that an association site of species $i$ is bound to species $j$. \textcolor{black}{Simply put, these association probabilities are the coordination numbers over the functionalities, i.e., the actual number of associating species over the maximal number.} Therefore, the volume fraction of free alkali cations can be written as $\phi_{10} = \phi_{+}(1 - p_{+-})^{f_{+}}$ and free anions as $\phi_{01} = \phi_-(1 - p_{-+})^{f_-}$\textcolor{black}{, assuming independent associations}. 

The association probabilities can be determined through the mass action law 
\begin{align}
\lambda \zeta = \dfrac{p_{+-}p_{-+}}{(1-p_{+-})(1-p_{-+})},
\label{eq:MAL}
\end{align}

\noindent and conservation of associations
\begin{align}
p_{+-}\psi_{+}=p_{-+}\psi_- = \zeta,
\label{eq:cons}
\end{align} 

\noindent where $\psi_{i}=f_{i}\phi_{i}/\xi_{i}$ is the total dimensionless concentration of association sites (per lattice)~\cite{mceldrew2020theory,McEldrewsalt2021}. Equations \eqref{eq:cons} and \eqref{eq:MAL} permit an explicit solution for the probabilities
\begin{align}
\lambda\zeta=\frac{1 + \lambda \psi_{\pm} - \sqrt{\left[1 + \lambda \psi_{\pm} \right]^2-4 \lambda^2 \psi_- \psi_{+}} }{2 }.
\label{eq:p}
\end{align}

\noindent where $\psi_{\pm} = \psi_+ + \psi_-$. It has been shown, from a bond percolation analysis, that the percolating ionic network onsets when $p_{+-}p_{-+} = [(f_{+}-1)(f_--1)]^{-1}$, which we also call the gel-point (see Refs.~\citenum{mceldrew2020theory,McEldrewsalt2021}). To determine sol and gel properties in the post-gel regime, we use Flory's treatment~\cite{mceldrew2020theory}.

\textcolor{black}{With the introduced association probabilities, the cluster distribution mainly depends on the value of the association constant, $\lambda$. This can be taken as a free parameter, but $\lambda$ can also be constrained from MD simulations or Raman/IR measurements from using Eqs.~\eqref{eq:MAL}-\eqref{eq:cons}, which can be inferred from the coordination numbers~\cite{Phelan2025Effect}. For this to be possible, the functionalities of each species also needs to be known. Again, this can be directly inferred from MD simulations, and so these are also not free fitting parameters~\cite{mceldrew2020corr,mceldrew2021ion,McEldrewsalt2021,Goodwin2023}. The only other parameters of the model are the volumes of species, but again these can be determined from simple calculations~\cite{mceldrew2020corr,mceldrew2021ion,McEldrewsalt2021,Goodwin2023}. If the free energy of forming a single association, $\Delta f = \Delta u - T\Delta s$, and the regular solution term, $\chi$, need to be determined, the composition of the SiILs must be varied to extract $\lambda$'s composition dependence~\cite{McEldrewsalt2021}. Therefore, while this is a phenomenological theory with parameters to tune for specific chemistries, none of these parameters are free, and so the predictions of the theory should be robust, provided its assumptions are being adhered to. The main assumption of the theory that can be broken is the Cayley-tree approximation. In SiILs, loop formation has been found at equilibrium, which causes disagreements between the theory and MD simulations~\cite{McEldrewsalt2021,Zhang2024}.}

\subsection{Reversible Aggregation Kinetics}

Next we outline the non-equilibrium theory for reversible ionic aggregation kinetics in bulk SiILs. This has not been explored in the context of concentrated electrolytes before. However, in the context of patchy-particle systems, an ostensibly similar system, the kinetics of aggregation has been well studied~\cite{Sciortino2011CO,Corezzi2012,Corezzi2010,Bianchi2007,Corezzi2009,Sciortino2008,Tartaglia2009}. In this section, the presented formalism closely follows to the works of Sciortino, Tartaglia and co-workers~\cite{Sciortino2011CO,Corezzi2012,Corezzi2010,Bianchi2007,Corezzi2009,Sciortino2008,Tartaglia2009}, and Dongen and Ernst~\cite{Dongen1984}. \textcolor{black}{First we write down the considered aggregation equilibrium which are the terms that go into the reversible Smoluchowski equation. The remainder of this section then specifies the rate kernels of these reversible aggregation and fragmentation processes. }

The reversible Smoluchowski aggregation equation governs the time evolution of thermoreversible clustering. This master equation describes how the change in concentration of an aggregate of rank $lm$, in the two-component case studied here, with concentration $c_{lm}(t)$, depends on the rates of formation and destruction of this aggregate
\begin{equation}
    c_{l-l'm-m'} + c_{l'm'}  \ce{<=>[K^{(f)}_{l,l',m,m'}][F^{(f)}_{l,l',m,m'}]} c_{lm},
    \label{eq:req1}
\end{equation}
\begin{equation} 
    c_{lm} + c_{l'm'}  \ce{<=>[K^{(r)}_{l,l',m,m'}][F^{(r)}_{l,l',m,m'}]} c_{l+l'm+m'}.
    \label{eq:req2}
\end{equation}

\noindent The first reversible aggregation process, Eq.~\eqref{eq:req1}, describes the rate of formation of $c_{lm}$ through the binding between $c_{l-l',m-m'}$ and $c_{l'm'}$ with a rate of $K^{(f)}_{l,l',m,m'}$~\cite{Tartaglia2009,Dongen1984}. Note we have left the rate constant to generally depend on the values of $l$, $l'$, $m$ and $m'$, with exactly how the rate depends on these values being shown later. In this same equilibrium, $c_{lm}$ can be removed through breaking an association to form $c_{l-l'm-m'}$ and $c_{l'm'}$ with rate $F^{(f)}_{l,l',m,m'}$. While the second reversible aggregation reaction, Eq.~\eqref{eq:req2}, describes the removal of $c_{lm}$ through the formation of an association with $c_{l',m'}$ to form $c_{l+l',m+m'}$, with a rate of $K^{(r)}_{l,l',m,m'}$, again generally given. In addition, there is the formation of $c_{lm}$ from the dissociation of $c_{l+l'm+m'}$, with rate $F^{(r)}_{l,l',m,m'}$~\cite{Tartaglia2009,Dongen1984}. These reversible aggregation reactions can be seen in the reversible Smoluchowski aggregation equation
\begin{multline}
    \dfrac{d c_{lm}}{dt} = \sum^{\infty}_{\substack{l'=0,\\ m'=0}}\left(c_{l+l'm+m'}F^{(r)} - c_{lm}c_{l'm'}K^{(r)}\right),
    \\ + \dfrac{1}{2}\sum_{\substack{l'+l'' = l,\\ m'+m''=m}}\left(c_{l'm'}c_{l''m''}K^{(f)} - c_{lm}F^{(f)}\right)
\label{eq:SM_ME}
\end{multline}

\noindent where the subscripts of $K^{(f/r)}_{l,l',m,m'}$ and $F^{(f/r)}_{l,l',m,m'}$ were dropped for clarity, and the explicit time-dependence of the concentrations suppressed~\cite{Tartaglia2009}.

\textcolor{black}{To solve the reversible Smoluchowski aggregation equation, we now must specify the rate kernels of the processes. First we can introduce the elementary rates for the formation of a single association, $k$, and the rate of a single dissociation, $f$, which is related to the association constant at equilibrium through $\lambda = k/f$~\cite{Tartaglia2009,Dongen1984}. Following Refs.~\citenum{Tartaglia2009,Dongen1984}, the coalescence rate for two aggregates is related to the number of dangling association sites in each aggregate and the individual rate of forming associations. For a cluster of rank $lm$ combining with a cluster of rank $l'm'$, this is}
\begin{equation}
    K^{(r)}_{l,l',m,m'} = k\sigma_{lm}\sigma_{l'm'},
\end{equation}

\noindent \textcolor{black}{where, for example}
\begin{equation}
    \sigma_{lm} = (f_{+}l-l-m+1)!(f_{-}m-m-l+1)!.
\end{equation}

\noindent \textcolor{black}{For a free cation and free anion associating, the rate of this process is not simply $k$, but $K^{(r)}_{1,0,0,1} = kf_+!f_-!$. }

At equilibrium ($t = \infty$), the steady state solution of Eq.~\eqref{eq:req2} imposes~\cite{Tartaglia2009}
\begin{equation}
    c_{lm}(\infty)c_{l'm'}(\infty)K^{(r)} = F^{(r)}c_{l+l'm+m'}(\infty).
    \label{eq:infeq}
\end{equation}

\noindent Therefore, from Eq.~\eqref{eq:infeq}, the full fragmentation rate is related to the rate of an individual fragmentation through
\begin{equation}
    F^{(r)}_{l,l',m,m'} =  f\sigma_{lm}\sigma_{l'm'}\dfrac{W_{lm}W_{l'm'}}{W_{l+l',m+m'}}.
\end{equation}

\noindent The same procedure can be repeated for Eq.~\eqref{eq:req1}, although not shown here~\cite{Tartaglia2009,Dongen1984}.

In principle, this determines the reversible Smoluchowski aggregation equation to be solved~\cite{Tartaglia2009,Dongen1984}. Directly solving the reversible Smoluchowski aggregation equation is a formidable task, however. Even in the irreversible Smoluchowski equation, direct solutions are limited to simple initial conditions and certain choices of the rate kernels. However, Flory and Stockmayer, and later others, demonstrated that the association probabilities could be used as the central variable, greatly simplifying the problem~\cite{Tartaglia2009,Dongen1984}. 

Instead, the time dependence of the association probabilities is solved for, considering the addition and removal of single associations through this ensemble averaged quantity. It was shown that these time-dependent association probabilities used in the equilibrium cluster distribution, Eq.~\eqref{eq:clust}, is a solution of the reversible Smoluchowski aggregation equation, in a quasi-equilibrium way~\cite{Tartaglia2009,Dongen1984}. In what follows, we solve the time-dependence of the association probabilities under different approximations. \textcolor{black}{In the main text, we solve a simpler ``symmetric case'' where there is only one association probability~\cite{Tartaglia2009,Dongen1984}, and in the SI the more general ``asymmetric'' case of two association probabilities is shown.}

\subsection{Symmetric case}

Typically, the functionalities of the alkali metal cation and anion are different in SiILs, which generally means their association probabilities are not equal~\cite{McEldrewsalt2021}. \textcolor{black}{For} the association probabilities to be equal, we must have $\psi_+ = \psi_-$ (from the conservation of associations), which can (sometimes) be achieved at a specific composition in SiILs. Since $f_+ = 4$ and $f_- = 3$~\cite{Zhang2024}, to achieve $p_{+-} = p_{-+}$, there must be $3/4$ times the number of alkali metal cations as anions. In this section, we consider $p_{+-} = p_{-+} = p$ generally and do not specify exact values of $f_i$ or $\phi_i$ explicitly, other than they satisfy $\psi_+ = \psi_- = \psi$ to ensure a single association probability. 

As demonstrated in Refs.~\citenum{Tartaglia2009,Dongen1984}, the time-dependence of $p(t)$ is determined from the macroscopic rate equation
\begin{equation}
\dfrac{d [A^R]}{d t} = k[A^F]^2 - f[A^R]
\label{eq:MRA_sym}
\end{equation}

\noindent where $[A^R] = \psi p$ is the concentrated of bound association sites, and $[A^F] = \psi (1 - p)$ is the concentration of unbound association sites. The rate of forming a single association, $k$, pre-multiplies $[A^F]^2$, to obtain the overall rate of forming associations. While the rate of breaking a single association, $f$, pre-multiplies $[A^R]$, to obtain the overall rate of breaking associations~\cite{Tartaglia2009,Dongen1984}. Similar to the equilibrium theory, this macroscopic rate equation assumes independence and equal reactivities of each association. Thus, the first-order differential which describes the time-dependence of the association probability is given by 
\begin{equation}
\dfrac{1}{k}\dfrac{d p}{d t} = \psi (1 - p)^2 - \dfrac{p}{\lambda}.
\label{eq:ptge_sym}
\end{equation}

\noindent In the SI, we show how this equation can be derived from considering a limiting case of ion pair formation~\cite{Goodwin2022Kornyshev}. In the limit of long-time, $dp(\infty)/dt = 0$, we find
\begin{equation}
\psi\lambda = \dfrac{p(\infty)}{[1 - p(\infty)]^2},
\label{eq:mass_sym}
\end{equation}

\noindent which is simply the mass action law, as in Eq.~\eqref{eq:MAL}, where the conservation of associations has been applied. Since the elemental rates, and therefore, the association constant are determined by the final state, the equilibrium state is naturally incorporated into the differential equation. Before there was a step-change in conditions that results in a variation of the associations, different rate constants and equilibrium constants applied. Upon the step-change, these rates are forgotten, but they initially persist through the initial values of the association probabilities~\cite{Tartaglia2009,Dongen1984}.

Applying the initial condition of $p(0) = p^0$, we find the solution to be~\cite{Tartaglia2009,Dongen1984}
\begin{equation}
p(t) = \dfrac{p^{\infty}(1 - \Lambda e^{-t/\tau})}{1 - p^{\infty}p^{\infty}\Lambda e^{-t/\tau}}
\label{eq:pt_sym}
\end{equation}

\noindent where $p(\infty) = p^\infty$ has been introduced, and we have defined the following parameters
\begin{equation}
\Lambda = \dfrac{p^{\infty} - p^0}{p^{\infty}(1 - p^{\infty}p^0)},
\label{eq:L_sym}
\end{equation}

\noindent which is a statement of the direction that the association probabilities are changing (if they are increasing or decreasing) and by how much, and 
\begin{equation}
\tau^{-1} = \dfrac{k}{\lambda}\sqrt{\left[1 + 2\lambda \psi \right]^2-4 \lambda^2 \psi^2} = k\psi\dfrac{1 - p^{\infty}p^{\infty}}{p^{\infty}},
\label{eq:tau_sym}
\end{equation}

\noindent is the inverse time-scale that controls the evolution of the association probabilities~\cite{Tartaglia2009,Dongen1984}. We find that this inverse relaxation time depends on two main factors. Firstly, it is proportional to the rate of forming an association~\cite{Tartaglia2009,Dongen1984}. Therefore, different electrolytes could be expected to have different relaxation times, based on this absolute scale, which is information that is removed in the association constant, since that depends on the ratio of the rates~\cite{mceldrew2021ion,molinari2019general,molinari2019transport}. The second term is proportional to a function of the probabilities, which cause a divergent time scale when the association probabilities approach 1. Therefore, in principle, extremely long relaxation times of electrolytes could be obtained from reversible aggregation. \textcolor{black}{Note that even though there are many values of the rate kernels from the different aggregates, these all collapse into a single exponential with one time constant, $\tau$. }

The solution of $p(t)$, in Eq.~\eqref{eq:pt_sym}, then completely determines the changes in the aggregates in the system, which has been shown to be a solution of the reversible Smoluchowski aggregation equation. Therefore, the aggregates change in a quasi-equilibrium way, i.e., they always have a distribution that obeys an equilibrium cluster distribution [Eq.~\eqref{eq:clust} with the association constant which changes in time, calculated using the association probabilities], with their changes being completely described through $p(t)$. This was noted by Sciortino and co-workers~\cite{Sciortino2011CO,Corezzi2012,Corezzi2009} \textcolor{black}{as a connection between} time and temperature through $p$, i.e., there is a time in the non-equilibrium aggregation when $p(t) = p^\infty(T)$, which can be thought \textcolor{black}{as} the temperature changing \textcolor{black}{with time} throughout the simulation. 

In the post-gel regime, Eq.~\eqref{eq:pt_sym} still applies, but the total association probabilities can be decomposed into sol and gel contributions, using Flory's post-gel treatment~\cite{mceldrew2020theory,McEldrewsalt2021}. For more information on how the reversible Smoluchowski aggregation equation is modified, \textcolor{black}{see} Ref.~\citenum{Dongen1984}. 

We can also study various limits of Eq.~\eqref{eq:ptge_sym}. An example is irreversible aggregation, $k >> f$ and $\lambda = \infty$, described by
\begin{equation}
\dfrac{1}{k}\dfrac{d p}{d t} = \psi (1 - p)^2.
\label{eq:ptge_sym_irr_agg}
\end{equation}

\noindent Starting with an initial condition of $p(0) = 0$, the solution to this equation is $p(t) = k\psi t/(1 + k\psi t)$. This can be seen to be a limit of Eq.~\eqref{eq:pt_sym} from taking $p(0) = 0$ with $\lambda \approx \infty$, i.e., $p^\infty \approx 1$. 

If, by contrast, we study irreversible fragmentation, where the initial aggregates can only break apart, we would have 
\begin{equation}
\dfrac{d p}{d t} = -fp.
\label{eq:ptge_sym_irr_frag}
\end{equation}

\noindent The solution of this is simply $p(t) = p^0e^{-ft}$. Therefore, the aggregates should exponentially decay, with the rate being solely determined by the rate of breaking a single association. This can be seen to be a limit of Eqs.~\eqref{eq:pt_sym}-\eqref{eq:tau_sym}, through taking $p^\infty$ to be zero and $\lambda = 0$. Commonly, the lifetime of associations (residence times) are calculated from MD simulations, and this differential equation would describe it, if there was a single time-scale~\cite{molinari2019general,Self2019,Fong2019}.

\textcolor{black}{In this section and the previous one, it should be noted that the dynamical version of this theory introduces a new parameter, namely the rate of forming $k$ (or breaking $f$) an association. Only a single parameter is introduced, since the rate of breaking (forming) an association is related through the association constant. This parameter needs to be determined in the theory from comparison against MD simulations or experiments. In principle, it could be extracted from residence times of equilibrium simulations, and applied to the non-equilibrium case, or directly fit from simulations/experiments.}

\section{Molecular Dynamics}

To test the equations \textcolor{black}{derived} in the Theory Section, we perform atomistic, classical molecular dynamics simulations of an example SiIL. We choose to study NaTFSI in EMIMTFSI. To perform these simulations, we use LAMMPS~\cite{Thompson2022} and the CL\&P force field~\cite{lopes2012} with charges rescaled by 0.7 to mimic the effect of polarization~\cite{molinari2019general,molinari2019transport}. The initial inputs were generated with fftool and packmol~\cite{martinez2009packmol}. We studied mole fractions of $x=0.25,0.5,2/3,0.75$ in NaTFSI$_x$EMIMTFSI$_{1-x}$, with 400 TFSI$^-$ anions in the simulation cell and varying numbers of each cation. A timestep of 1~fs was utilized, with all hydrogen atoms fixed by shake.

To equilibrate the system, we initially performed a temperature annealing cycle. Starting from a 1~ns run at 300~K (for example\textcolor{black}{, as different temperatures are investigated, this is the final temperature is initiated here}), we ramped the system to 450~K over 1~ns, and held at that temperature for another 1~ns, before the same procedure is repeated with a final temperature of 600~K (for higher temperature simulations, e.g., 400~K and 500~K, the same change in temperatures were performed for the annealing cycle). An equivalent cooling cycling was then applied. Then another 10~ns equilibration was performed \textcolor{black}{at the target temperature}. This was performed in NPT, where we use a temperature thermostat with 100~fs, and 1000~fs for the pressure. In NVT, at the equilibrium density, we then equilibrated for another 10~ns. 

To model the effect of the \textcolor{black}{non-equilibrium} kinetics of aggregation, we need to step-change a property\textcolor{black}{/variable} in the simulation. The most natural way of \textcolor{black}{introducing non-equilibrium conditions} is to step-change the temperature. Several simulations were performed, where, for example, we first equilibrated the electrolyte with a density that corresponds to 300~K \textcolor{black}{(the final, target temperature)} at a temperature of 500~K, which causes the associations to break from the higher temperature. Then we would step change the temperature to 300~K, i.e., the temperature of the density that the box is equilibrated to in NVT. This, however, only produces very modest changes in the coordination numbers. The results of these simulations are \textcolor{black}{shown} in the SI, but will be discussed in the main text.

To produce a more substantial effect, to investigate more thoroughly if the kinetic equations derived perform well, we investigated a more artificial test. In MD simulations, we have control over the interactions in the electrolytes. Therefore, we can turn on and off the electrostatic interactions, which is a main driving force for associations in the electrolyte. After the NVT equilibration \textcolor{black}{at the target temperature with all interactions}, as described before, we rescaled all charges by 10$^6$, making them essentially 0, and equilibrated the electrolyte for 1~ns. Then the original charges (0.7 of the CL\&P force field~\cite{lopes2012}) of the atoms were reinstated, and the changes in coordination numbers tracked in time. Note that turning on and off the charges pumps/removes significant energy into the system. Therefore, while tracking how coordination numbers change in time, we more aggressively thermostated the electrolyte (2~fs for 1000 steps, then 5~fs for 1000 steps, and 10~fs for 1000 steps before the normal value was used for the remainder). This ensured the temperature remained approximately constant (if this was not performed, temperatures in the \textcolor{black}{range of} 10$^3$~K were found). Note we did not find this aggressive thermostating qualitatively changed the results, with only a small quantitative effect \textcolor{black}{observed}. 

The association probabilities defined in Eq.~\eqref{eq:p} are directly linked to the average coordination number of alkali cations by anions ($CN_{+-} = f_{+}p_{+-}$), as well the coordination number of anions by alkali cations ($CN_{-+} = f_-p_{-+}$). From the MD simulations, we define if a Na$^+$ and TFSI$^-$ are associated if the O of TFSI$^-$ is within 3.3\AA of Na$^+$, or vice versa~\cite{molinari2019general,molinari2019transport,McEldrewsalt2021}. This was chosen based on the first time $g(r)$ reaches 1 after the first peak~\cite{mceldrew2021ion}, see SI for details. Other real-space cutoffs were investigated too. While the exact quantitative numbers change, the overall conclusions found here remain independent of this choice. \textcolor{black}{Other methods for defining associations exist, such as kinetic criteria~\cite{feng2019free}, but as there are large energy and dynamical changes occurring in the non-equilibrium simulations, the real-space method appears to be the most robust. } Note that if more than 1 O in TFSI$^-$ is within the cutoff to the same Na$^+$, this just counts as 1 association, i.e., bidendate associations only count as a single association here, as we are counting based on species. To \textcolor{black}{quantify} the ionic aggregates in the simulations, \textcolor{black}{i.e., the number of ions in an aggregate and the number of associations within that aggregate,} we construct an adjacency matrix \textcolor{black}{of the ions}. 

To determine a persistence/residence time (shown in the SI), we use a Heaviside step function between Na$^+$ and O in TFSI$^-$ to count associations (so $H(r,t) = 1$ if $r \geq r_{cut}$ and $H(r,t) = 0$ if $r < r_{cut}$) to determine an autocorrelation function $\langle H(r,t)H(r,0) \rangle$, which can be used to extract a persistence/residence time~\cite{molinari2019general,Self2019,Fong2019,feng2019free}.

\section{Results}

\textcolor{black}{In this section, the results from the MD simulations are shown and used to validate the derived equations from the Theory Section. This is a self-contained section and does not require the Theory Section to understand the main messages found. We describe how the coordination numbers change with time after the step-change is applied, then the percolation criteria is analyzed and the cluster distribution is used to confirm the existence of ionic networks for the symmetric case. In the SI additional results are shown for the asymmetric case, further supporting the observations shown in the main text.}

\subsection{Symmetric case}

\begin{figure}
    \centering
    \includegraphics[width=0.95\linewidth]{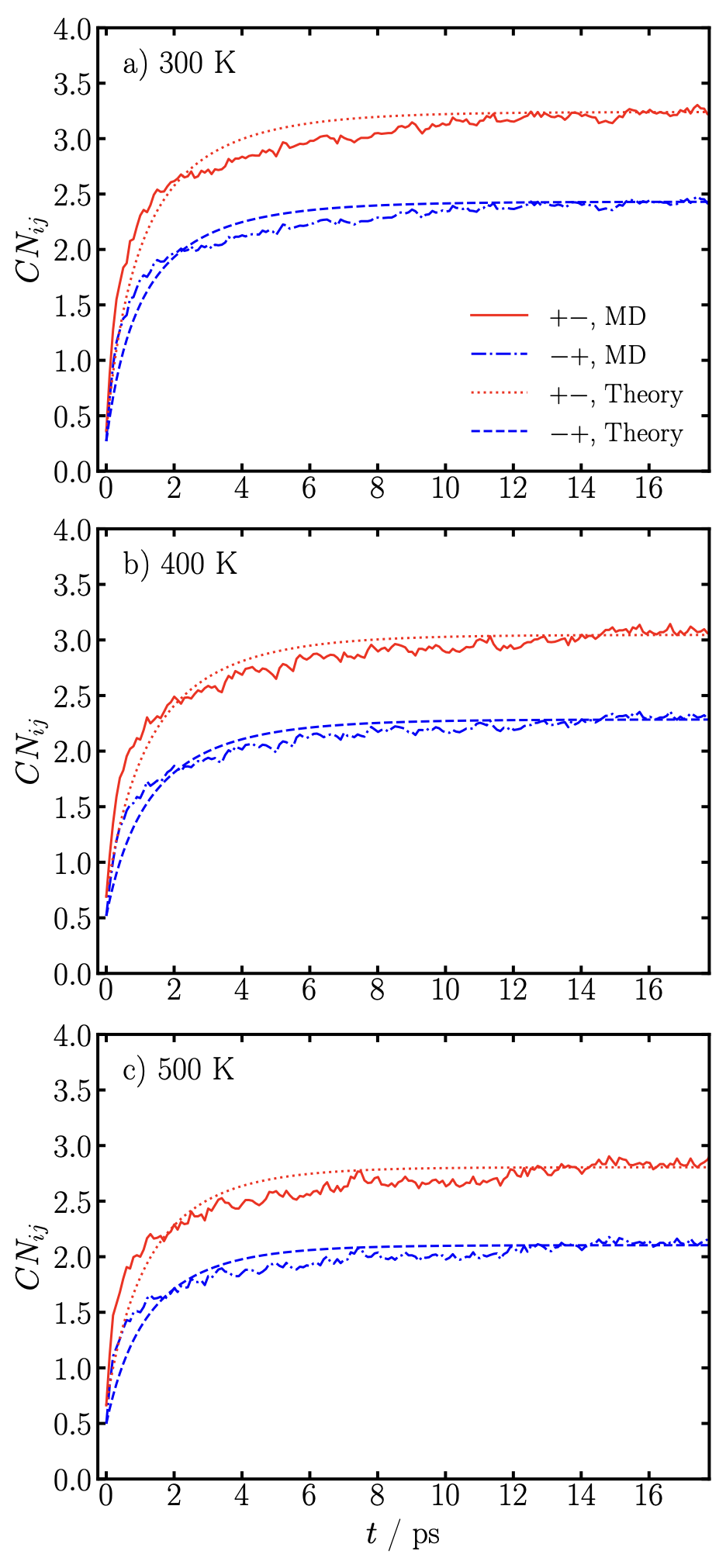}
    \caption{Coordination numbers between Na$^+$ and TFSI$^-$ in NaTFSI$_{0.75}$EMIMTFSI$_{0.25}$ as a function of time, for the charge rescaling case, at the indicated temperatures, from MD simulations and theory. }
    \label{fig:CN_sym}
\end{figure}

In Fig.~\ref{fig:CN_sym} we show how the coordination numbers between Na$^+$ and TFSI$^-$ change with time from the non-equilibrium MD simulations, with a comparison against \textcolor{black}{the developed} theory. Here we study the symmetric case of the SiIL NaTFSI$_{0.75}$EMIMTFSI$_{0.25}$, since the functionality \textcolor{black}{(maximum number of associations it can form)} of Na$^+$ is 4 and TFSI$^-$ is 3, this electrolyte mixture has $p_{+-}(t) = p_{-+}(t)$ \textcolor{black}{(i.e., the probability of cations associating to anions, $p_{+-}$, becomes equal to the probability of anions associating to cations, $p_{-+}$)}. Remember, the coordination numbers are related to the association probabilities through $CN_{ij} = f_ip_{ij}$~\cite{mceldrew2021ion,McEldrewsalt2021,Goodwin2023}. We choose to show the results in terms of coordination numbers, as these are readily extracted from MD simulations, and it gives distinct numerical values for Na$^+$ and TFSI$^-$ in this symmetric case. The results for the simulations in Fig.~\ref{fig:CN_sym} correspond to the charge rescaling case (see Molecular Dynamics Section for further description). In the SI, we show results for the temperature rescaling non-equilibrium conditions. The same conclusions are drawn from those, but where smaller changes in coordination numbers are observed.

Figure~\ref{fig:CN_sym}a) shows the simulations at 300~K. Initially, even though the charges of all atoms are essentially 0, we find coordination numbers of $\sim$0.3. This can be attributed to the fact that a real-space cutoff has been used, and the volume has been constrained to the equilibrium value at 300~K with the normal values of the charge. We find that $CN_{+-}$ (number of TFSI$^-$ anions bound to Na$^+$ cations) and $CN_{-+}$ (number of Na$^+$ cations bound to TFSI$^-$ anions) initially increases rapidly over the first 2~ps of the simulation, after which they increase less rapidly. After 15~ps, these coordination numbers have \textit{approximately} saturated at their equilibrium values. 

To compare against the theory, we have to extract the initial and final association probabilities. Then the only unknown in the theory is the time-constant of the decay in the probabilities, $\tau$, since the rates of forming and breaking associations ($k$ and $f$, respectively) are not known. Using the MD data, we fitted $CN_{+-/-+}$ over the 17~ps displayed to extract $\tau$, which we report in Tab.~\ref{tab:symmetric}. We find that the fitted $\tau$ for each curve ($CN_{+-}$ and $CN_{-+}$) gives the same decay rate of $\sim$0.4~ps$^{-1}$. This corresponds to a decay time of $\sim$2.5~ps, which agrees well with the time-scale observed in Fig.~\ref{fig:CN_sym}a). Using this fitted decay constant, we plot the theory prediction in Fig.~\ref{fig:CN_sym}a), where reasonable agreement is found. Moreover, using Eq.~\eqref{eq:tau_sym}, we use this information to extract the rate constant of forming associations, $k$, which we also report in Tab.~\ref{tab:symmetric}. We find a value of 3~ps$^{-1}$ for the rate of forming associations, which is larger than the fitted inverse time constant and can be attributed to the rate of forming associations needing to be faster than the observed inverse rate constant, since they are reversible associations.

In Fig.~\ref{fig:CN_sym}b)\&c), we show analogous non-equilibrium coordination numbers as a function of time for 400~k and 500~k, respectively. Again fitting these (using the initial and final association probabilities from MD) provides the time constants $\sim$0.41 and $\sim$0.51~ps$^{-1}$, for 400~K and 500~K, respectively; and we find that the fitted decay constant for the theory results is a reasonable fit to the MD simulations. With increasing temperature, we observe the inverse time constant to increase, indicating the time constant decreases with temperature. Intuitively, this makes sense, since at higher temperatures the associations will be weaker, which should therefore take less time to form. In Tab.~\ref{tab:symmetric}, we also report the association constant ($\lambda$) at the studied temperatures, where we find it decreases with temperature. Again, using Eq.~\eqref{eq:tau_sym} we extract the rate constant of forming an association, $k$, and find $\sim$2.4 and $\sim$2.2~ps$^{-1}$, for 400~K and 500~K, respectively. These are summarized in Tab.~\ref{tab:symmetric}. With increasing temperature, we find $k$ decreases. If the process was activated, $\propto e^{-\beta E_a}$, it should increase with temperature. This suggests that the activation barrier, if present, is not large, and other factors are dominating the rate of forming an association. 

\begin{table}[]
\centering
\begin{tabular}{lccc}
\hline
    $T$ / K  & $\tau^{-1}$ / ps$^{-1}$  & $k$ / ps$^{-1}$  & $\lambda$ \\
    \hline
    300     &  0.401 &  3.000 & 71.178 \\
    400     &  0.418  &  2.398  & 42.383 \\
    500     &  0.511  &  2.231  & 24.880 \\
    \hline
\end{tabular}
\caption{Summary of inverse time scales ($\tau^{-1}$), rate of forming an association ($k$), and association constant ($\lambda$) for the studied temperatures of the symmetric case.}
\label{tab:symmetric}
\end{table}

While the theory also has an initial rapid increase over $\sim$2~ps, and slower increase after, the quantitative agreement with the MD simulations is not exact (same conclusion is obtained from the temperature rescaling simulations, see SI). We find that the theory under-predicts the MD simulations at short-times, but over-predicts the values of the coordination numbers at long-times. The only variable that could cause this disagreement is the decay constant ($\tau^{-1}$), or more specifically the rate constants of forming ($k$) and breaking ($f$) an association. Since a single value of this rate of forming an association does not fit the data well, it suggests that there is more than one time scale in the simulations. 

To investigate this further, we calculated the decay of the associations over time of, which can be used to extract residence/persistence times (see Molecular Dynamics Methods for the description). In other words, \textcolor{black}{after} an initial time in an equilibrium simulation, we counted the number of associations that remained after some given time. This is similar to the irreversible fragmentation solution discussed in the symmetric case. However, from the simulations, if those exact associations between a specific cation and anion reform, this causes the measure to increase, which would not be possible in the irreversible aggregation case. However, since species tend to diffuse and associate with other ions after dissociating, this measure generally decays with time. In the SI, we show this result for NaTFSI$_{0.75}$EMIMTFSI$_{0.25}$ at 300~K. What we find is that initially the associations decay by 10-20\% very rapidly over several picoseconds. After which, the decay of the initial associations becomes more gradual, and only changes over a $\sim$~ns time scale. Similar results were found by Molinari \textit{et al.} for SiILs~\cite{molinari2019transport}, Self and Fong in the context of battery electrolytes~\cite{Self2019,Fong2019}, and by Feng \textit{et al.} in the context of ILs~\cite{feng2019free}. Therefore, there are clearly multiple time-scales for the kinetics of aggregate formation in SiILs\textcolor{black}{, and the theory assuming these is one timescale is a limitation}.

\begin{figure}
    \centering
    \includegraphics[width=0.95\linewidth]{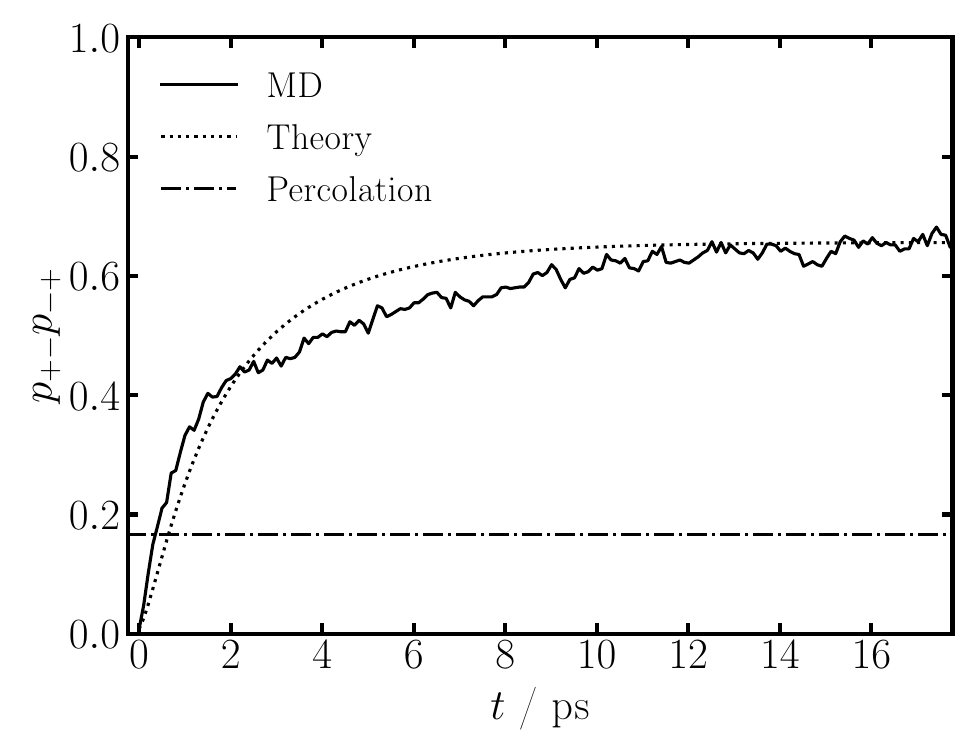}
    \caption{Gelation criteria in NaTFSI$_{0.75}$EMIMTFSI$_{0.25}$ as a function of time, for the charge rescaling case, at 300~K, for MD and theory. The percolation point is given by $p_{+-}p_{-+} = [(f_+ - 1)(f_- - 1)]^{-1}$.}
    \label{fig:perc_sym}
\end{figure}

\begin{figure*}
    \centering
    \includegraphics[width=0.95\linewidth]{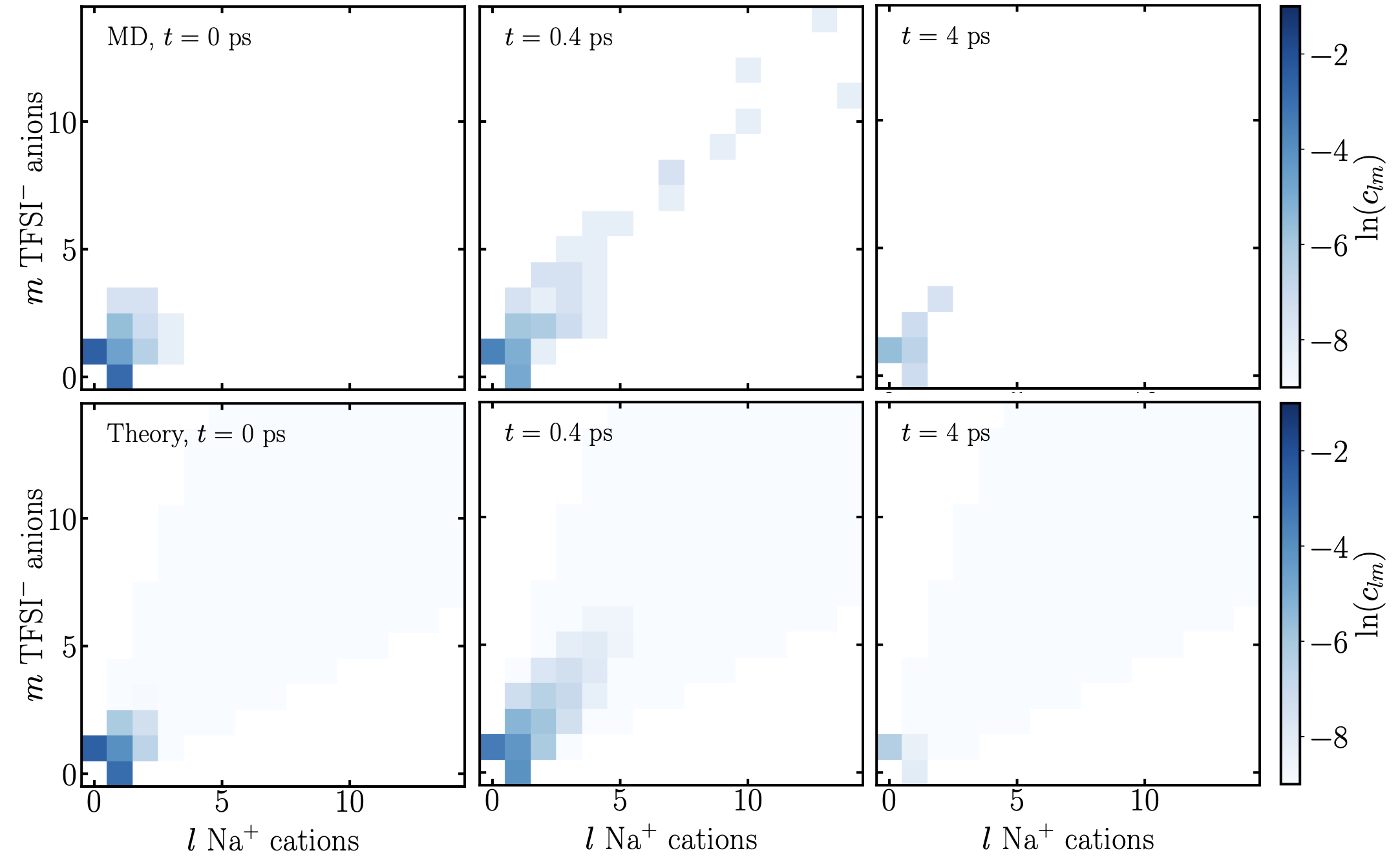}
    \caption{Cluster distributions $c_{lm}$, with $m$ TFSI$^-$ anions and $l$ Na$^+$ cations in an aggregate, at various times in the MD simulation and theory, as indicated, at 300~K.}
    \label{fig:cluster_sym}
\end{figure*}

To test this further, we fitted the coordination numbers over time, but only for first 5~ps. In the SI we show these fits, where much better agreement is found, since it is described by a single, fast association-forming time scale ($\tau^{-1} \approx 1$~ps$^{-1}$\textcolor{black}{, which gives $k \approx 4.73$~ps$^{-1}$}). Moreover, we also show that the coordination numbers increase slightly from the values in Fig.~\ref{fig:CN_sym}, over 10s of ps time scales. Therefore, in the MD simulations, the coordination numbers are initially controlled by the fast rate constant of forming associations of species that are already in proximity to each other, but at longer time scales this rate constant becomes limited by the diffusion/rearrangement of the electrolyte (not associations, since they should already be accounted for) to allow for the equilibrium associations to be reached~\cite{Self2019,Fong2019}. This explains why the theory underestimating the MD are short time scales, but overestimating the MD at long time scales. These observations pose a fundamental limitation of the theory, which assumes one rate constant, and motivates its development further. While the theory is not a perfect quantitative fit, it still provides insight into the underlying dynamics, and we further investigate it.

\textcolor{black}{Thus far, we have systemically investigated the best possible agreement between the theory and simulations from fitting the decay constants. While we have found that to obtain a good quantitative fit, development of the theory to have multiple time scales is required, it is regardless unsatisfactory to have free fitting parameters in the theory. All that needs to be determined for a parameter-free comparison is the rate of forming/breaking an association $k/f$. Since we have found multiple time-scales, these rates need to be determined at short and long times. From the irreversible aggregation solution and the residence/persistence calculations, we can assume that an approximation to breaking an association ($f$) is given by the inverse residence/persistence time, which can be converted to the rate of forming an association through $k = f\lambda$. For short times, we find $k_s \approx 5.88$ ps$^{-1}$ (given $\lambda = 26.73$), and at long-times we find $k_l \approx 0.03$ ps$^{-1}$ (from $\lambda = 163.02$). At short-times, the extracted rate of forming associations is close to the fitted value, previously discussed. Moreover, in the SI, we show good agreement between the theory with this calculated rate ($k_s$) over short-times and the MD simulations. We also show the fit for the long-time extracted $k_l$, but much worse agreement is found overall, since most of the change occurs at short-times, and there is a memory effect in these non-equilibrium conditions. The long-time behavior is improved, however, with this extracted value $k_l$. Therefore, overall, it appears that with development of the rate-kernels, a quantitative agreement without free fitting parameters is within reach.}

Using the time evolving association probabilities, we can \textcolor{black}{determine the} time a percolating ionic network, i.e., a gel, between Na$^+$ and TFSI$^-$ forms through tracking the product of the association probabilities, $p_{+-}p_{-+}$. When this value reaches $[(f_+ - 1)(f_- - 1)]^{-1}$, that is the critical criteria for bond percolation on a Bethe lattice, i.e., a Cayley tree~\cite{mceldrew2020theory,mceldrew2021ion}. In SiILs, the percolating ionic network is know to occur at $x\approx0.25$ from simulations~\cite{McEldrewsalt2021}; while the theory \textcolor{black}{often} predicts it at smaller mole fractions \textcolor{black}{at equilibrium}, owing to the formation of loops in the clusters~\cite{McEldrewsalt2021,Zhang2024}. 

In Fig.~\ref{fig:perc_sym} we show $p_{+-}p_{-+}$ as a function of time for NaTFSI$_{0.75}$EMIMTFSI$_{0.25}$ at 300~K. From the MD simulations, we find that the critical criteria for form a gel is already satisfied after $\sim$0.5~ps. The theory predicts the percolating ionic network to form at $\sim$1~ps, but this is because the theory is underestimating the association probabilities at short times \textcolor{black}{from the fitted decay constants}.

To check if the critical gel criteria is actually being satisfied in the simulations, we can inspect the cluster distribution~\cite{mceldrew2020theory,mceldrew2021ion,McEldrewsalt2021}. If we are in a pre-gel regime, we only expect small clusters (relative to the simulation size) to occur. As we approach the gel-point, the concentration of larger clusters continuously grows. When the gel-point is reached, there is simultaneously aggregates of all sizes \textcolor{black}{(within the simulation size)} and an aggregate that percolates throughout the simulation cell. In the post-gel regime, larger aggregates bind to the gel, increasing the gels volume fraction, leaving only small clusters. See Refs.~\citenum{mceldrew2020theory,mceldrew2021ion,McEldrewsalt2021} for more information. 

In Fig.~\ref{fig:cluster_sym}, we show the cluster distribution computed from MD simulations and theory at 3 times. At $t=0$, we find that free anions and cations dominate, with a smaller number of ion pairs, and even fewer aggregates containing a handful of ions. At $t=0.4$~ps, right at the critical gel time, we find that free anions still dominate the cluster distribution (as they usually do for SiILs~\cite{McEldrewsalt2021,Zhang2024}), but there are aggregates which exist with over 30 ions. At this time, we also find a single aggregate with a significant portion of the ions in the simulation cell. At later times than $t=0.4$~ps, the size of the aggregates decreases steadily. We show an example at $t=4$~ps, where free anions still dominate, and there are at most 5 ions in an aggregate, which is smaller than when the charges were scaled to 0 at $t=0$~ps. Therefore, the critical gel-time of the simulation predicted in Fig.~\ref{fig:perc_sym} corresponds well to the behavior of the aggregates in the simulation, shown in Fig.~\ref{fig:cluster_sym}, confirming the bond percolation criteria~\cite{mceldrew2020theory,mceldrew2021ion,McEldrewsalt2021}. The theory predictions (calculated with the cluster distribution, and the association probabilities and association constant changing in time) match the simulated ones qualitatively very well. Previously, for SiILs, as discussed in Ref.~\citenum{McEldrewsalt2021}, the theory cluster distribution did not agree well with that of the MD simulations, owing to the presence of loops in the aggregates~\cite{Zhang2024}. In the SI, we computed the cluster bond density, which quantifies how many loops exist in the aggregates. Unlike the equilibrium case~\cite{Zhang2024}, we only find a few loops, which could be why good agreement is obtained here.

\section{Discussion}

To summarize, we developed the kinetics of aggregation formalism of Sciortino, Tartaglia and co-workers~\cite{Sciortino2011CO,Sciortino2009PF,Corezzi2012,Corezzi2010,Bianchi2007,Corezzi2009,Sciortino2008,Tartaglia2009} from patchy particle systems to work for concentrated electrolytes, extending the equilibrium formalism developed by McEldrew \textit{et al.}~\cite{mceldrew2020theory,mceldrew2020corr,mceldrew2021ion,McEldrewsalt2021,Goodwin2023,Phelan2025Effect}. Comparisons to molecular dynamics simulations found good qualitative agreement, but quantitative differences were observed. These were investigated further, and found to result from multiple time scales existing in the concentrated electrolyte~\cite{molinari2019transport,Self2019,Fong2019}. Specifically, there is a fast time at short-times, corresponding to associations forming between species that are \textcolor{black}{already} in close proximity. After which the time scale slows, being more limited by diffusion and rearrangements of ions~\cite{Self2019}. While the theory is not quantitative, it does provide \textcolor{black}{analytical expressions with} insight into the overall mechanisms occurring and a framework to understand them. For example, we also demonstrated that the quasi-equilibrium solution applies to SiILs, where the cluster distribution changes in time from the changes in association probabilities, with this quasi-equilibrium solution connecting time and temperature through the association probabilities, since it can be thought that $p(t)$ is changing through equilibrium states of $p(T)$~\cite{Sciortino2011CO,Sciortino2009PF,Corezzi2012,Corezzi2010,Bianchi2007,Corezzi2009,Sciortino2008,Tartaglia2009}.

\textcolor{black}{The analysis here has been completely theoretical and computational based, but this does not limit the applicability of the developed formalism. In fact, as analytical expression for coordination numbers changing in time after a step-change in conditions, this directly lends itself to facile experimental and computational comparison. In battery electrolytes, Raman and IR spectroscopic methods are commonly employed to extract information about the coordination environments~\cite{Phelan2025Effect}, since there are often distinct features in the spectra for bound/unbound functional groups. If a step-change in temperature can be introduced into these electrolytes, and the response experimentally characterized, the developed equations can be used to obtain the rates of forming/breaking associations in these electrolytes directly from experiments. In addition, for computational work, the formalism presented here provides a framework to understand equilibrium and non-equilibrium changes in coordination environments and clusters, which can also be linked to other physiochemical properties~\cite{McEldrewsalt2021}.}

Using \textcolor{black}{the developed formalism and performed MD simulations}, we extracted the rate for the formation of an association, which was found to \textcolor{black}{change with time (from the short-time to long-time response) and,} vary slightly with temperature and composition of the electrolyte \textcolor{black}{(see SI)}. \textcolor{black}{We found that the short-time rate of forming associations is of the order of 5~ps$^{-1}$, which corresponds to a typical time for ions to rattle in the cages that form in concentrated electrolytes~\cite{feng2019free}. Whereas, the long time scale corresponds to time scales of 10s of ps, which is closer to the scale of diffusive time scales in concentrated electrolytes. Therefore, the magnitudes of the rates appear to be physically sensible~\cite{molinari2019general,molinari2019transport}. } As discussed in detail in Refs.~\citenum{Tartaglia2009,Dongen1984}, the rates of fragmentation ($F$) and coalescence ($K$) depend on the time-scale for an association to form and the diffusive time-scales of the species which associate\textcolor{black}{, which we also found here through the multiple time scales in the electrolytes. This transition from ``bond'' forming time-scale to diffusion time-scale can be modeled with rate kernels which depend on the association probabilities, allowing for transitions from rates limited by ``bond'' forming to diffusion~\cite{Corezzi2012}. These observations are interesting, as multiple time scales in simulations have already been observed~\cite{molinari2019general,molinari2019transport}, but if we have a distribution of clusters, it might be expected that there are a distribution of relaxation times. In the formalism developed here, however, this does not emerge, but one (or two) time scales emerge from the theory as this cluster distribution is accounted for in an ensemble averaged way through the association probabilities which change with time.}

In the context of the temperature dependence, an exponential law \textcolor{black}{would} be expected for an activated process, \textcolor{black}{meaning that} the rate \textcolor{black}{of forming associations would} increase with increasing temperature. Since we do not observe this, it suggests this process is not an activated one. While in the context of different compositions \textcolor{black}{(see SI)}, changes in dielectric environment and viscosity might also be expected to change these rates substantially. However, we also do not observe strong effects from these variables. \textcolor{black}{Both of these observations could be a result of the short-time scale dominating the effect, where there is limited influence from temperature/composition on forming association between ions that are already in proximity.} It is known that residence times of different SiILs~\cite{molinari2019transport}, ILs~\cite{feng2019free}, WiSEs~\cite{mceldrew2021ion} and battery electrolytes~\cite{Self2019} can \textcolor{black}{be quite different}, which suggests different rates of forming associations\textcolor{black}{, but this is likely dominated by the diffusive time-scale}. Future work should investigate these observations in more depth.

\textcolor{black}{While we have mainly extracted the rate constants from fitting the MD simulations, it was also demonstrated that approximate rate values can be obtained from bulk, equilibrium calculations of residence/persistence. The obtained rates at short-times agreed well with the MD simulation results at short times, demonstrating a parameter-free model could be obtained. However, the limitation of the multiple time scales prevented a full comparison here, with developments into the rate kernels changing throughout the processes being a promising direction for a parameter-free quantitative match~\cite{Corezzi2012}.}

\textcolor{black}{Another source of error between the theory and simulations could be a result of the employed force field. A simple non-polarizable force-field was utilized here, with scaled charges to model the effects of using a polarizable force field~\cite{molinari2019general,molinari2019transport}. If another force-field is utilized for the simulations, the non-equilibrium dynamics would almost certainly change, and this could change the agreement between the theory and simulations observed here. Moreover, comparisons to \textit{ab initio} MD and machine learning interatomic potential simulations could also be readily achieved through step-changing temperature. It is not \textit{a priori} clear, however, if changing the force field will improve the agreement with the theory. On point is clear, however, that the time and length scales that are simulated will limit the resulting rates of forming/breaking associations, while experimental measurements are macroscopic so the finite size effects should not play a role in bulk measurements.}

\textcolor{black}{Here we have investigated correlations between a select cation and anion ($+-$), using a real-space cutoff to define associations between species. This formalism is more general, however. It can be applied to understand like-charge correlations too, albeit being much weaker. In the SI, we show how the ``coordination numbers'' between like charges ($++$ and $--$) change with time. We find that the $++$ correlations only have a fast, sub-diffusive time scale and the theory is able to reproduce these dynamics well. However, the $--$ correlations have a non-monotonic dependence, initially reducing before returning to slightly above the initial value, and the theory is unable to describe its behaviour. Interestingly, this suggests that the anion plays a more major role in the multiple time scales of the electrolyte than the alkali metal cations. Beyond this, the requirement of associations between species (cation-anion) can be relaxed, so taking into account bidentate associations is possible, or higher-order encapsulating associations~\cite{molinari2020chelation}, for example. Additionally, other methods for defining the associations, such as kinetic-based cutoffs~\cite{feng2019free}, could be used to investigate the non-equilibrium effects studied here.}

\textcolor{black}{Moreover,} from the electrolytes that have studied with this formalism so far, SiILs is one of the most challenging~\cite{McEldrewsalt2021,Zhang2024}. Its \textcolor{black}{highly} concentrated and correlated nature pushes the boundaries of the theory, where loops within the aggregates form (breaking the Cayley tree approximation), making it more difficult to apply the theory~\cite{McEldrewsalt2021,Zhang2024}. While SiILs are the most challenging example that have been applied \textcolor{black}{the equilibrium} theory to so far, we nonetheless \textcolor{black}{found that the non-equilibrium formalism could capture well the dynamics of SiILs}. In future work, applied to other electrolytes, we expect the developed formalism to perform even better.

It is hoped that this provides the first link between a microscopic theory of ionic associations and non-equilibrium changes to electrolyte structure, which is further built upon. In what follows, we outline possible extensions and connections to other areas in concentrated electrolytes\textcolor{black}{, highlighting where experimental verification can enter}. While this is not an exhaustive list, we hope it provides inspiration for what might now be possible.

In the context of ionic transport in concentrated electrolytes, there is simultaneously vehicular motion of the aggregates/clusters and structural diffusion~\cite{Self2019,Fong2019}. To determine the dominant mechanism, Self and Fong \textit{et al.}~\cite{Self2019,Fong2019} suggested comparing length scales of the aggregates ($L^s$) to the length scales over which they move in a vehicular way. The later is determined from $L^c = \sqrt{6D_i\tau_{res}}$, where $D_i$ is the self diffusion coefficient and $\tau_{res}$ is the residence time of the association. If $L^c > L^s$, there is vehicular dominated motion, while if $L^c < L^s$ it is structural dominated motion. There are clearly multiple time scales for associations in electrolytes, as found here for SiILs, and by others~\cite{molinari2019transport,feng2019free,mceldrew2021ion,Self2019,Fong2019}. Self and Fong \textit{et al.}~\cite{Self2019,Fong2019} suggested using the long-time scale, and ignoring the short, sub-diffusive changes in associations. If the short time scale is used, which is attributed to the binding/breaking of associations in proximity, this would suggests structural diffusion dominates more. While if the long time scale is used, it suggest more of a vehicular motion could be dominant. The results here suggest that this short-time scale is important, indicating that transport in these electrolytes could be more dominated by structural diffusion. Using the formalism developed here, it might be possible to more quantitatively link vehicular and structural diffusion modes. 

As discussed in the Results and Theory Sections, concentrated electrolytes have been predicted to form a percolating ionic network, i.e., a gel phase~\cite{mceldrew2020theory,mceldrew2020corr,mceldrew2021ion,McEldrewsalt2021} (to date, only one experimental claim of a gel from associations has been reported~\cite{reber2021anion}). The terminology of the gel phase comes from the inherited language from polymer systems~\cite{flory1942thermodynamics,stockmayer1943theory,stockmayer1944theory,stockmayer1952molecular,ishida1997,tanaka1989,tanaka1990thermodynamic,tanaka1994,tanaka1995,tanaka1998,tanaka1999,tanaka2002}. In thermoreversible polymers the time scales of the associations are much longer than electrolytes, since the associations \textcolor{black}{(``bonds'')} are much stronger. Upon gelation, this gives rise to a divergence in viscosity and the onset of an elastic response~\cite{mceldrew2020theory}. It is not clear if the elastic response will be measurable from the shorter time scales of the associations in concentrated electrolytes, however. In the context of experimental measurements, non-Newtonian behavior of WiSEs has been \textcolor{black}{observed}~\cite{Yamaguchi2024nn}, suggesting that in WiSE these associations could be long-lived enough. Significantly more work is required in this area, from theory, simulations and experiments, to understand how similar electrolytes are to polymers in this regard, i.e., how good the analogy between ionic associations and thermoreversible bonds. Using the developed formalism, it should be possible to include the finite time scales of associations, to more quantitatively predict the frequency dependence of the viscoelastic response of the fluid~\cite{Tanaka2023,Tanaka2024}.

Here we have only studied bulk electrolytes, but the technologies which use these electrolytes often require contact with electrified interfaces~\cite{Fedorov2014,Gebbie2023,Zheng2017Uni,Cheng2022Sol}. How concentrated electrolytes accumulate at charged interfaces, i.e., the electrical double layers (EDL), has been studied at equilibrium in the context of McEldrew \textit{et al.}'s formalism for ILs~\cite{Goodwin2022EDL,Goodwin2022Kornyshev}, SiILs~\cite{markiewitz2024,Zhang2024} and WiSEs~\cite{Markiewitz2025}. Using the formalism developed here, the charging dynamics of the EDL could be studied with the relaxation of the aggregates. \textcolor{black}{Experimentally, this could be achieved with time-resolved SIERAS or WAXS~\cite{Han2020}, which would also allow for the changes in coordination environments to be probed.} While the aggregation relaxation time scale appears to be faster than the electrolyte transport, studying the relaxation of aggregates in confinement could result in interesting interplay with the transport of ions. Building on this, the desolvation dynamics of battery electrolytes at the electrodes would be another area to investigate, to more quantitatively determine the time scales involving how the solvationc shell is striped from the active cation as it intercalates into the electrode~\cite{Rubinstein2026}. As a first step, the removal of one solvent from the diffuse EDL to Helmholtz layer could be investigated~\cite{GoodwinHelm2025}, building all the way up to the complete removal and intercalation of the active ion.

Finally, in the context of ``anomalous underscreening''~\cite{Gebbie2013,Gebbie2015,Han2020,Han2021WiSE,groves2021surface,smith2016electrostatic,perez2017underscreening,Espinosa2023rev}, it has been shown that with slower approaches of the interfaces the long-ranged interactions vanish~\cite{Cross2026}. Therefore, this effect is \textcolor{black}{likely} not from bulk electrostatic interactions, but a large part is hydrodynamic interactions~\cite{Cross2026,Zhang2024}. It remains to be seen if it can be completely described by classical fluid mechanics, however. In confinement with these concentrated electrolytes, the ionic aggregation effects could also be playing a role~\cite{Zhang2024}. In the context of SiILs, Zhang \textit{et al.}~\cite{Zhang2024} found force separation, extremely long force decay lengths, and changes in refractive indices of the electrolyte between mica. This made them conclude there was confinement induced changes in electrolyte composition, possibly leading to the formation of a gel~\cite{Zhang2024}. Further developing the formalism here to confined geometries, also with changing composition, could provide insight into the role of ionic aggregation kinetic in these measurements. We have found that the time scale of ionic aggregation $\tau$ can diverge if the association probabilities approach 1, which could give rise to extremely long relaxation times. However, the short-time scale of association appears to be of the order of $\sim$ps (which could also arise from the finite size of the simulation cells). Further work should explore if the aggregate relaxations could help explain these puzzling non-equilibrium measurements.

\section{Acknowledgments}

I am extremely grateful for the support and stimulating discussion from M. McEldrew, A. Kornyshev, C. Phelan and R. Espinosa-Marzal. I also acknowledge discussions with J. P. de Souza, R. Weatherup, N. Molinari, M. Markiewitz and M. Bazant. Z.A.H.G acknowledges support through the Glasstone Research Fellowship in Materials and The Queen's College, University of Oxford.

\bibliography{WiSE}

@article{Zaccone2013,
  title={Colloidal gelation with variable attraction energy},
  author={Alessio Zaccone and J\'er^{o}me J. Crassous and Matthias Ballauff},
  journal={J. Chem. Phys.},
  volume={138},
  pages={104908},
  year={2013},
}

@article{Zaccone2011,
  title={Quantifying the Reversible Association of Thermosensitive Nanoparticles},
  author={Alessio Zaccone and Jerome J. Crassous and Benjamin B\'eri and Matthias Ballauff},
  journal={Phys. Rev. Lett.},
  volume={107},
  pages={168303},
  year={2011},
}

@article{Rubinstein2026,
  title={Finite-time dehydration as a chemical reaction and its effect on counterion selectivity},
  author={I Rubinstein and MZ Bazant and NA Kononenko and VV Nikonenko and B Zaltzman},
  journal={Physical Review E},
  volume={113},
  pages={045404},
  year={2026},
}

@article{Thompson2022,
  title={LAMMPS - a flexible simulation tool for particle-based materials modeling at the atomic, meso, and continuum scales},
  author={A. P. Thompson and H. M. Aktulga and R. Berger and D. S. Bolintineanu and W. M. Brown and P. S. Crozier and P. J. in 't Veld and A. Kohlmeyer and S. G. Moore and T. D. Nguyen and R. Shan and M. J. Stevens and J. Tranchida and C. Trott and S. J. Plimpton},
  journal={Comp Phys Comm},
  volume={271},
  pages={10817},
  year={2022},
}

@article{Fong2019,
  title={Ion Transport and the True Transference Number in Nonaqueous Polyelectrolyte Solutions for Lithium Ion Batteries},
  author={Kara D. Fong and Julian Self and Kyle M. Diederichsen and Brandon M. Wood and Bryan D. McCloskey and Kristin A. Persson},
  journal={ACS Cent. Sci.},
  volume={5},
  pages={2019},
  year={1250--1260},
}

@article{Self2019,
  title={Transport in Superconcentrated LiPF$_6$ and LiBF$_4$/Propylene Carbonate Electrolytes},
  author={Julian Self and Kara D. Fong and Kristin A. Persson},
  journal={ACS Energy Lett.},
  volume={4},
  pages={2843--2849},
  year={2019},
}

@article{Tian2022,
  title={Electrolyte Solvation Structure Design for Sodium Ion Batteries},
  author={Zhengnan Tian and Yeguo Zou and Gang Liu and Yizhou Wang and Jian Yin and Jun Ming and Husam N. Alshareef},
  journal={Adv. Sci.},
  volume={9},
  pages={2201207},
  year={2022},
}

@article{Xu2021desol,
  title={Li$^+$-solvation/desolvation dictates interphasial processes on graphitic anode in Li ion cells},
  author={Kang Xu and Arthur von Wald Cresce},
  journal={J. Mater. Res., Vol. , No. 18, Sep 28, },
  volume={27},
  pages={2327},
  year={2012},
}

@article{Shirley2022,
  title={Designing better electrolytes},
  author={Y. Shirley Meng and Venkat Srinivasan and Kang Xu},
  journal={Science},
  volume={378},
  pages={1065},
  year={2022},
}

@article{Yamaguchi2024nn,
  title={Non-Newtonian Dynamics in Water-in-Salt Electrolytes},
  author={Tsuyoshi Yamaguchi and Andrei Dukhin and Young-Jay Ryu and Dongzhou Zhang and Oleg Borodin and Miguel A. Gonz\`alez and Osamu Yamamuro and David L. Price and Marie-Louise Saboungi},
  journal={J. Phys. Chem. Lett.},
  volume={15},
  pages={76--80},
  year={2024},
}

@article{Tanaka2024,
  title={Viscoelastic Properties of Stimuli-Responsive Transient Polymer Networks},
  author={Fumihiko Tanaka},
  journal={Macromolecules},
  volume={57},
  pages={10600--10614},
  year={2024},
}

@article{groschel2013,
  title={Guided hierarchical co-assembly of soft patchy nanoparticles},
  author={Andr\'{e}H. Gr\"{o}schel and Andreas Walther and Tina I. L\"{o}bling and Felix H. Schacher and Holger Schmalz and Axel H. E. M\"{u}ller},
  journal={Nature},
  volume={503},
  pages={247},
  year={2013},
}

@article{Umar2013,
  title={Aggregation Behavior of Oppositely Charged Gold Nanorods in Aqueous Solution},
  author={Aminah Umar and Sung-Min Choi},
  journal={J. Phys. Chem. C},
  volume={117},
  pages={11738--11743},
  year={2013},
}

@article{Tartaglia2009,
  title={Gel formation through reversible and irreversible aggregation},
  author={Piero Tartaglia},
  journal={J. Phys.: Condens. Matter},
  volume={21},
  pages={504109},
  year={2009},
}

@article{Sciortino2008,
  title={Growth of equilibrium polymers under non-equilibrium conditions},
  author={Francesco Sciortino and Cristiano De Michele and Jack F Douglas},
  journal={J. Phys.: Condens. Matter},
  volume={20},
  pages={155101},
  year={2008},
}

@article{Corezzi2009,
  title={Connecting Irreversible to Reversible Aggregation: Time and Temperature},
  author={S. Corezzi and C. De Michele and E. Zaccarelli and P. Tartaglia and F. Sciortino},
  journal={J. Phys. Chem. B},
  volume={113},
  pages={1233--1236},
  year={2009},
}

@article{Bianchi2007,
  title={Fully Solvable Equilibrium Self-Assembly Process: Fine-Tuning the Clusters Size and the Connectivity in Patchy Particle Systems},
  author={Emanuela Bianchi and Piero Tartaglia and Emilia La Nave and Francesco Sciortino},
  journal={J. Phys. Chem. B},
  volume={111},
  pages={11765--11769},
  year={2007},
}

@article{Tanaka2023,
  title={Gelation Time of Network-Forming Polymer Solutions with Reversible Cross-Link Junctions of Variable Multiplicity},
  author={Fumihiko Tanaka},
  journal={Gels},
  volume={9},
  pages={379},
  year={2023},
}

@article{Corezzi2012,
  title={Chemical and physical aggregation of smallfunctionality particles},
  author={Silvia Corezzi and Daniele Fiorettoac and Francesco Sciortino},
  journal={Soft Matter},
  volume={8},
  pages={11207--11216},
  year={2012},
}

@article{Corezzi2010,
  title={Modeling the Crossover between Chemically and Diffusion-Controlled Irreversible Aggregation in a Small-Functionality Gel-Forming System},
  author={S. Corezzi and D. Fioretto and C. De Michele and E. Zaccarelli and F. Sciortino},
  journal={J. Phys. Chem. B},
  volume={114},
  pages={3769--3775},
  year={2010},
}

@article{Dongen1984,
  title={Kinetics of Reversible Polymerization},
  author={P. G. J. van Dongen and M. H. Ernst},
  journal={Journal of Statistical Physics},
  volume={37},
  pages={301},
  year={1984},
}

@article{Sciortino2009PF,
  title={A parameter-free description of the kinetics of formation of loop-less branched structures and gels},
  author={Francesco Sciortino and Cristiano De Michele and Silvia Corezzi and John Russo and Emanuela Zaccarelli and Piero Tartaglia},
  journal={Soft Matter},
  volume={5},
  pages={2571--2575},
  year={2009},
}

@article{Sciortino2011CO,
  title={Reversible gels of patchy particles},
  author={Francesco Sciortino and Emanuela Zaccarelli},
  journal={Current Opinion in Solid State and Materials Science},
  volume={15},
  pages={246--253},
  year={2011},
}

@article{Cross2026,
  title={Short-range electrostatic screening in ionic liquids as inferred
by direct force measurements},
  author={Benjamin Cross and Leo Garcia and Elisabeth Charlaix and Patrick Kekicheff},
  journal={PNAS},
  volume={123},
  pages={e2517939123},
  year={2026},
}

@article{Gebbie2023,
  title={Linking Electric Double Layer Formation to Electrocatalytic Activity},
  author={Matthew A. Gebbie and Beichen Liu and Wenxiao Guo and Seth R. Anderson and Samuel G. Johnstone},
  journal={ACS Catal.},
  volume={13},
  pages={16222--16239},
  year={2023},
}

@article{Qisheng23JACS,
  title={Effect of the Electric Double Layer (EDL) in Multicomponent Electrolyte Reduction and Solid Electrolyte Interphase (SEI) Formation in Lithium Batteries},
  author={Qisheng Wu and Matthew T. McDowell and Yue Qi},
  journal={JACS},
  volume={145},
  pages={2473--2484},
  year={2023},
}

@article{Cheol2023HE,
  title={High-entropy electrolytes for practical lithium metal batteries},
  author={Sang Cheol Kim and Jingyang Wang and Rong Xu and Pu Zhang and Yuelang Chen and Zhuojun Huang and Yufei Yang and Zhiao Yu and Solomon T. Oyakhire and Wenbo Zhang and Louisa C. Greenburg and Mun Sek Kim and David T. Boyle and Philaphon Sayavong and Yusheng Ye and Jian Qin and Zhenan Bao and Yi Cui },
  journal={Nature Energy},
  volume={8},
  pages={814--826},
  year={2023},
}

@phdthesis{McEldrewPhD,
    author = {Michael McEldrew},
    title = {Ion Aggregation, Correlated Ion Transport and the Double Layer in Super-Concentrated Electrolytes},
    school = {MIT},
    year = {2021}
}

@article{Tkachenko2025,
  title={Dehydration-Driven Ion Aggregation and the Onset of Gelation in ZnCl2 Solution},
  author={Alexei V. Tkachenko and Chuntian Cao and Amy C. Marschilok and Deyu Lu},
  journal={arXiv:2512.18167},
  volume={},
  pages={},
  year={2025},
}

@article{Phelan2025Effect,
  title={The Effect of Salt Concentration on Electrochemical Stability in Li-ion Battery Electrolytes},
  author={Conor Phelan and Arihant Bhandari and Jasper Singh and Michael Fraser nad Erik Bj\"orklund and Jack Swallow and Peter Bencok and Clare Grey and Zachary Goodwin and Chris Skylaris and Robert Weatherup},
  journal={https://chemrxiv.org/doi/pdf/10.26434/chemrxiv-2025-k903g},
  volume={},
  pages={},
  year={2025},
}

@article{GoodwinHelm2025,
  title={Theory of Cation Solvation in the Helmholtz Layer of Li-Ion Battery Electrolytes},
  author={Zachary AH Goodwin and Daniel M Markiewitz and Qisheng Wu and Yue Qi and Martin Z Bazant},
  journal={ACS Applied Energy Materials},
  volume={8},
  pages={8376--8387},
  year={2025},
}

@article{Markiewitz2025,
  title={Ionic associations and hydration in the electrical double layer of water-in-salt electrolytes},
  author={Daniel M Markiewitz and Zachary AH Goodwin and Qianlu Zheng and Michael McEldrew and Rosa M Espinosa-Marzal and Martin Z Bazant},
  journal={ACS Applied Materials \& Interfaces},
  volume={17},
  pages={29515-29534},
  year={2025},
}

@article{Han2021WiSE,
  title={Nanoheterogeneity of LiTFSI Solutions Transitions Close to a Surface and with Concentration},
  author={Mengwei Han and Ruixian Zhang and Andrew A. Gewirth and Rosa M. Espinosa-Marzal},
  journal={Nano Lett.},
  volume={21},
  pages={2304--2309},
  year={2021},
}

@article{Pedro2020,
  title={Interfacial layering in the electrical double layer of ionic liquids},
  author={J Pedro de Souza and Zachary AH Goodwin and Michael McEldrew and Alexei A Kornyshev and Martin Z Bazant},
  journal={Phys. Rev. Lett.},
  volume={125},
  pages={116001},
  year={2020},
}

@article{Han2020,
  title={Insight into the Electrical Double Layer of Ionic Liquids Revealed
through Its Temporal Evolution},
  author={M Han and H Kim and C Leal and M Negrito and J D Batteas and R M Espinosa-Marzal},
  journal={Adv Mater Interfaces},
  volume={7},
  pages={2001313},
  year={2020},
}

@inproceedings{Espinosa2023rev,
  title={Colloidal Interactions in Ionic Liquids—The Electrical Double Layer Inferred from Ion Layering and Aggregation},
  author={Espinosa-Marzal, Rosa M and Goodwin, Zachary AH and Zhang, Xuhui and Zheng, Qianlu},
  booktitle={One Hundred Years of Colloid Symposia: Looking Back and Looking Forward},
  pages={123--148},
  year={2023},
  organization={ACS Publications}
}

@article{Zhou2022Agg,
  title={Beyond Local Solvation Structure: Nanometric Aggregates in Battery Electrolytes and Their Effect on Electrolyte Properties},
  author={Zhou Yu and Nitash P. Balsara and Oleg Borodin and Andrew A. Gewirth and Nathan T. Hahn and Edward J. Maginn and Kristin A. Persson and Venkat Srinivasan and Michael F. Toney and Kang Xu and Kevin R. Zavadil and Larry A. Curtiss and Lei Cheng},
  journal={ACS Energy Lett.},
  volume={7},
  pages={461--470},
  year={2022},
}

@article{Chagas2019,
  title={Ionic Liquid-Based Electrolytes for Sodium-Ion Batteries: Tuning Properties To Enhance the Electrochemical Performance of Manganese-Based Layered Oxide Cathode},
  author={Luciana Gomes Chagas and Sangsik Jeong and Ivana Hasa and Stefano Passerini},
  journal={ACS Appl. Mater. Interfaces},
  volume={11},
  pages={22278--22289},
  year={2019},
}

@article{Goodwin2022Kornyshev,
  title={Cracking Ion Pairs in the Electrical Double Layer of Ionic Liquids},
  author={Zachary AH Goodwin and Alexei A Kornyshev},
  journal={Electrochim. Acta},
  volume={434},
  pages={141163},
  year={2022},
}

@article{Goodwin2022EDL,
  title={Gelation, Clustering and Crowding in the Electrical Double Layer of Ionic Liquids},
  author={Zachary AH Goodwin and Michael McEldrew and J Pedro de Souza and Martin Z Bazant and Alexei A Kornyshev},
  journal={J. Chem. Phys.},
  volume={157},
  pages={094106},
  year={2022},
}

@article{Goodwin2023,
  title={Theory of Cation Solvation and Ionic Association in Nonaqueous Solvent Mixtures},
  author={Zachary A H Goodwin and Michael McEldrew and Boris Kozinsky and Martin Z Bazant},
  journal={PRX Energy},
  volume={2},
  pages={013007},
  year={2023},
}

@article{McEldrewsalt2021,
  title={Salt-in-ionic-liquid electrolytes: Ion network formation and negative effective charges of alkali metal cations},
  author={Michael McEldrew and Zachary A H Goodwin and Nicola Molinari and Boris Kozinsky and Alexei A Kornyshev and Martin Z Bazant},
  journal={J. Phys. Chem. B},
  volume={125},
  pages={13752--13766},
  year={2021},
}

@article{Welton2011,
  title={Room-Temperature Ionic Liquids: Solvents for Synthesis and Catalysis. 2},
  author={Jason P. Hallett and Tom Welton},
  journal={Chem. Rev.},
  volume={111},
  pages={3508–3576},
  year={2011},
}

@article{Welton1999,
  title={Room-Temperature Ionic Liquids. Solvents for Synthesis and Catalysis},
  author={Tomas Welton},
  journal={Chem. Rev.},
  volume={99},
  pages={2071–2084},
  year={1999},
}

@article{mceldrew2021ion,
  title={Ion Clusters and Networks in Water-in-Salt Electrolytes},
  author={McEldrew, Michael and Goodwin, Zachary AH and Bi, Sheng and Kornyshev, Alexei and Bazant, Martin Z},
  journal={J. Electrochem. Soc.},
  year={2021},
  volume = {168},
  pages = {050514},
  publisher={IOP Publishing}
}

@article{mceldrew2020corr,
  title={Correlated Ion Transport and the Gel Phase in Room Temperature Ionic Liquids},
  author={Michael McEldrew and Zachary A. H. Goodwin and Hongbo Zhao and Martin Z. Bazant and Alexei A. Kornyshev},
  journal={J. Phys. Chem B},
  volume={125},
  pages={2677–2689},
  year={2021},
}

@article{lourenco2021theoretical,
  title={Theoretical Investigation of the Na+ Transport Mechanism and the Performance of Ionic Liquid-Based Electrolytes in Sodium-Ion Batteries},
  author={Lourenco, Tuanan C and Dias, Luis Gustavo and Da Silva, Juarez LF},
  journal={ACS Appl. Energy Mater.},
  volume={4},
  number={5},
  pages={4444--4458},
  year={2021},
  publisher={ACS Publications}
}

@article{Lannelongue2018,
author = {Lannelongue, Pierre and Bouchal, Roza and Mourad, El{\'{e}}onore and Bodin, Charlotte and Olarte, Marco and le Vot, Steven and Favier, Fr{\'{e}}d{\'{e}}ric and Fontaine, Olivier},
doi = {10.1149/2.0951803jes},
issn = {0013-4651},
journal = {Journal of The Electrochemical Society},
month = {mar},
number = {3},
pages = {A657--A663},
publisher = {The Electrochemical Society},
title = {{“Water-in-Salt” for Supercapacitors: A Compromise between Voltage, Power Density, Energy Density and Stability}},
url = {http://jes.ecsdl.org/lookup/doi/10.1149/2.0951803jes},
volume = {165},
year = {2018}
}

@article{reber2021anion,
  title={Anion Selection Criteria for Water-in-Salt Electrolytes},
  author={Reber, David and Grissa, Rabeb and Becker, Maximilian and K{\"u}hnel, Ruben-Simon and Battaglia, Corsin},
  journal={Advanced Energy Materials},
  volume={11},
  number={5},
  pages={2002913},
  year={2021},
  publisher={Wiley Online Library}
}

@article{Gebbie2013,
author = {Gebbie, M. A. and Valtiner, M. and Banquy, X. and Fox, E. T. and Henderson, W. A. and Israelachvili, J. N.},
doi = {10.1073/pnas.1307871110},
issn = {0027-8424},
journal = {Proceedings of the National Academy of Sciences},
month = {jun},
number = {24},
pages = {9674--9679},
title = {{Ionic liquids behave as dilute electrolyte solutions}},
url = {http://www.pnas.org/cgi/doi/10.1073/pnas.1307871110},
volume = {110},
year = {2013}
}

@article{Gebbie2015,
author = {Gebbie, M. A. and Dobes, H. A. and Valtiner, M. and Israelachvili, J. N.},
journal = {Proceedings of the National Academy of Sciences},
pages = {7432–7437},
title = {{Long-range electrostatic screening in ionic liquids}},
volume = {112},
year = {2015}
}

@article{Kjellander2016,
author = {Kjellander, Roland},
doi = {10.1039/C6CP02418A},
issn = {1463-9076},
journal = {Physical Chemistry Chemical Physics},
number = {28},
pages = {18985--19000},
title = {{Decay behavior of screened electrostatic surface forces in ionic liquids: the vital role of non-local electrostatics}},
url = {http://xlink.rsc.org/?DOI=C6CP02418A},
volume = {18},
year = {2016}
}

@article{Bazant2011,
author = {Bazant, Martin Z and Storey, Brian D and Kornyshev, Alexei A},
title = {{Double Layer in Ionic Liquids: Overscreening versus Crowding}},
journal = {Phys. Rev. Lett.},
issn = {0031-9007},
month = {jan},
number = {4},
pages = {046102},
volume = {106},
year = {2011},
url = {https://link.aps.org/doi/10.1103/PhysRevLett.106.046102},
doi = {10.1103/PhysRevLett.106.046102}
}

@article{vatamanu2017,
  title={Ramifications of water-in-salt interfacial structure at charged electrodes for electrolyte electrochemical stability},
  author={Vatamanu, Jenel and Borodin, Oleg},
  journal={J. Phys. Chem. Lett.},
  volume={8},
  number={18},
  pages={4362--4367},
  year={2017},
  publisher={ACS Publications}
}

@article{Fedorov2014,
author = {Fedorov, Maxim V. and Kornyshev, Alexei A.},
doi = {10.1021/cr400374x},
issn = {0009-2665},
journal = {Chem. Rev.},
month = {mar},
number = {5},
pages = {2978--3036},
title = {{Ionic Liquids at Electrified Interfaces}},
url = {http://pubs.acs.org/doi/10.1021/cr400374x},
volume = {114},
year = {2014}
}

@article{Kornyshev2007,
author = {Kornyshev, Alexei A.},
doi = {10.1021/jp067857o},
issn = {1520-6106},
journal = {J. Phys. Chem. B},
month = {may},
number = {20},
pages = {5545--5557},
title = {{Double-Layer in Ionic Liquids:  Paradigm Change?}},
url = {http://pubs.acs.org/doi/abs/10.1021/jp067857o},
volume = {111},
year = {2007}
}

@article{Suo2015,
author = {Suo, Liumin and Borodin, Oleg and Gao, Tao and Olguin, Marco and Ho, Janet and Fan, Xiulin and Luo, Chao and Wang, Chunsheng and Xu, Kang},
doi = {10.1126/science.aab1595},
issn = {1095-9203},
journal = {Science},
month = {nov},
number = {6263},
pages = {938-43},
pmid = {26586759},
publisher = {American Association for the Advancement of Science},
title = {{``Water-in-salt" electrolyte enables high-voltage aqueous lithium-ion chemistries.}},
url = {http://www.ncbi.nlm.nih.gov/pubmed/26586759},
volume = {350},
year = {2015}
}

@article{Suo2013,
author = {Suo, Liumin and Hu, Yong-Sheng and Li, Hong and Armand, Michel and Chen, Liquan},
doi = {10.1038/ncomms2513},
issn = {2041-1723},
journal = {Nat. Commun.},
month = {dec},
number = {1},
pages = {1481},
publisher = {Nature Publishing Group},
title = {{A new class of Solvent-in-Salt electrolyte for high-energy rechargeable metallic lithium batteries}},
url = {http://www.nature.com/articles/ncomms2513},
volume = {4},
year = {2013}
}

@article{Wang2016,
author = {Wang, Jianhui and Yamada, Yuki and Sodeyama, Keitaro and Chiang, Ching Hua and Tateyama, Yoshitaka and Yamada, Atsuo},
doi = {10.1038/ncomms12032},
issn = {2041-1723},
journal = {Nat. Commun.},
month = {jun},
pages = {12032},
publisher = {Nature Publishing Group},
title = {{Superconcentrated electrolytes for a high-voltage lithium-ion battery}},
url = {http://www.nature.com/doifinder/10.1038/ncomms12032},
volume = {7},
year = {2016}
}

@article{Yamada2016,
author = {Yamada, Yuki and Usui, Kenji and Sodeyama, Keitaro and Ko, Seongjae and Tateyama, Yoshitaka and Yamada, Atsuo},
doi = {10.1038/nenergy.2016.129},
issn = {2058-7546},
journal = {Nat. Energy},
month = {aug},
number = {10},
pages = {16129},
publisher = {Nature Publishing Group},
title = {{Hydrate-melt electrolytes for high-energy-density aqueous batteries}},
url = {http://www.nature.com/articles/nenergy2016129},
volume = {1},
year = {2016}
}

@article{kuhnel2017,
  title={A high-voltage aqueous electrolyte for sodium-ion batteries},
  author={K{\"u}hnel, Ruben-Simon and Reber, David and Battaglia, Corsin},
  journal={ACS Energy Lett.},
  volume={2},
  number={9},
  pages={2005--2006},
  year={2017},
  publisher={ACS Publications}
}

@article{lopes2012,
  title={CL\&P: A generic and systematic force field for ionic liquids modeling},
  author={Lopes, Jos{\'e} N Canongia and P{\'a}dua, Ag{\'\i}lio AH},
  journal={Theoretical Chemistry Accounts},
  volume={131},
  number={3},
  pages={1129},
  year={2012},
  publisher={Springer}
}

@article{martinez2009packmol,
  title={PACKMOL: a package for building initial configurations for molecular dynamics simulations},
  author={Mart{\'\i}nez, Leandro and Andrade, Ricardo and Birgin, Ernesto G and Mart{\'\i}nez, Jos{\'e} Mario},
  journal={J. Comput. Chem.},
  volume={30},
  number={13},
  pages={2157--2164},
  year={2009},
  publisher={Wiley Online Library}
}

@article{xu2004nonaqueous,
  title={Nonaqueous liquid electrolytes for lithium-based rechargeable batteries},
  author={Xu, Kang},
  journal={Chem. Rev.},
  volume={104},
  number={10},
  pages={4303--4418},
  year={2004},
  publisher={ACS Publications}
}

@article{borodin2017liquid,
  title={Liquid structure with nano-heterogeneity promotes cationic transport in concentrated electrolytes},
  author={Borodin, Oleg and Suo, Liumin and Gobet, Mallory and Ren, Xiaoming and Wang, Fei and Faraone, Antonio and Peng, Jing and Olguin, Marco and Schroeder, Marshall and Ding, Michael S and others},
  journal={ACS nano},
  volume={11},
  number={10},
  pages={10462--10471},
  year={2017},
  publisher={ACS Publications}
}

@article{Borodin2017Mod,
  title={Modeling Insight into Battery Electrolyte Electrochemical Stability and Interfacial Structure},
  author={Oleg Borodin and Xiaoming Ren and Jenel Vatamanu and Arthur von Wald Cresce and Jaroslaw Knap and Kang Xu},
  journal={Acc. Chem. Res.},
  volume={50},
  pages={2886--2894},
  year={2017},
}

@article{Welton2025,
  title={A reflection on salts dissolved in salts: ionic liquid mixtures – ion pairs, ion pairing and ionic liquids},
  author={Tom Welton},
  journal={Chem. Sci.},
  volume={16},
  pages={18976--18984},
  year={2025},
}

@article{lim2018,
  title={Nanometric water channels in water-in-salt lithium ion battery electrolyte},
  author={Lim, Joonhyung and Park, Kwanghee and Lee, Hochan and Kim, Jungyu and Kwak, Kyungwon and Cho, Minhaeng},
  journal={Journal of the American Chemical Society},
  volume={140},
  number={46},
  pages={15661--15667},
  year={2018},
  publisher={ACS Publications}
}

@article{tanaka1989,
  title={Theory of thermoreversible gelation},
  author={Tanaka, Fumihiko},
  journal={Macromolecules},
  volume={22},
  number={4},
  pages={1988--1994},
  year={1989},
  publisher={ACS Publications}
}

@article{tanaka1994,
  title={Thermoreversible gelation with junctions of variable multiplicity},
  author={Tanaka, Fumihiko and Stockmayer, Walter H},
  journal={Macromolecules},
  volume={27},
  number={14},
  pages={3943--3954},
  year={1994},
  publisher={ACS Publications}
}

@article{ishida1997,
  title={Theoretical study of the postgel regime in thermoreversible gelation},
  author={Ishida, Masahiko and Tanaka, Fumihiko},
  journal={Macromolecules},
  volume={30},
  number={13},
  pages={3900--3909},
  year={1997},
  publisher={ACS Publications}
}

@article{tanaka1995,
  title={Thermoreversible gelation of hydrated polymers},
  author={Tanaka, Fumihiko and Ishida, Masahiko},
  journal={J. Chem. Soc. Faraday Trans.},
  volume={91},
  number={16},
  pages={2663--2670},
  year={1995},
  publisher={The Royal Society of Chemistry}
}

@article{tanaka1998,
  title={Thermoreversible gelation of associating polymers},
  author={Tanaka, Fumihiko},
  journal={Physica A: Statistical Mechanics and its Applications},
  volume={257},
  number={1-4},
  pages={245--255},
  year={1998},
  publisher={Elsevier}
}

@article{tanaka2002,
  title={Theoretical study of molecular association and thermoreversible gelation in polymers},
  author={Tanaka, Fumihiko},
  journal={Polym. J.},
  volume={34},
  number={7},
  pages={479},
  year={2002},
  publisher={Nature Publishing Group}
}

@article{tanaka1999,
  title={Thermoreversible gelation with two-component networks},
  author={Tanaka, Fumihiko and Ishida, Masahiko},
  journal={Macromolecules},
  volume={32},
  number={4},
  pages={1271--1283},
  year={1999},
  publisher={ACS Publications}
}

@book{flory1953principles,
  title={Principles of polymer chemistry},
  author={Flory, Paul J},
  year={1953},
  publisher={Cornell University Press}
}

@article{feng2019free,
  title={Free and Bound States of Ions in Ionic Liquids, Conductivity, and Underscreening Paradox},
  author={Feng, Guang and Chen, Ming and Bi, Sheng and Goodwin, Zachary AH and Postnikov, Eugene B and Brilliantov, Nikolai and Urbakh, Michael and Kornyshev, Alexei A},
  journal={Phys. Rev. X},
  volume={9},
  number={2},
  pages={021024},
  year={2019},
  publisher={APS}
}

@article{stockmayer1943theory,
  title={Theory of molecular size distribution and gel formation in branched-chain polymers},
  author={Stockmayer, Walter H},
  journal={The Journal of chemical physics},
  volume={11},
  number={2},
  pages={45--55},
  year={1943},
  publisher={AIP}
}

@article{stockmayer1944theory,
  title={Theory of molecular size distribution and gel formation in branched polymers II. General cross linking},
  author={Stockmayer, Walter H},
  journal={The Journal of Chemical Physics},
  volume={12},
  number={4},
  pages={125--131},
  year={1944},
  publisher={AIP}
}

@article{stockmayer1952molecular,
  title={Molecular distribution in condensation polymers},
  author={Stockmayer, Walter H},
  journal={J. Polym. Sci.},
  volume={9},
  number={1},
  pages={69--71},
  year={1952},
  publisher={Wiley Online Library}
}

@article{tanaka1990thermodynamic,
  title={Thermodynamic theory of network-forming polymer solutions. 1},
  author={Tanaka, Fumihiko},
  journal={Macromolecules},
  volume={23},
  number={16},
  pages={3784--3789},
  year={1990},
  publisher={ACS Publications}
}

@article{flory1942thermodynamics,
  title={Thermodynamics of high polymer solutions},
  author={Flory, Paul J},
  journal={The Journal of chemical physics},
  volume={10},
  number={1},
  pages={51--61},
  year={1942},
  publisher={AIP}
}

@article{molinari2019transport,
  title={Transport anomalies emerging from strong correlation in ionic liquid electrolytes},
  author={Molinari, Nicola and Mailoa, Jonathan P and Craig, Nathan and Christensen, Jake and Kozinsky, Boris},
  journal={J. Power Sources},
  volume={428},
  pages={27--36},
  year={2019},
  publisher={Elsevier}
}

@article{molinari2019general,
  title={General Trend of a Negative Li Effective Charge in Ionic Liquid Electrolytes},
  author={Molinari, Nicola and Mailoa, Jonathan P and Kozinsky, Boris},
  journal={J. Phys. Chem. Lett.},
  volume={10},
  number={10},
  pages={2313--2319},
  year={2019},
  publisher={ACS Publications}
}

@article{borodin2020uncharted,
  title={Uncharted Waters: Super-Concentrated Electrolytes},
  author={Borodin, Oleg and Self, Julian and Persson, Kristin A and Wang, Chunsheng and Xu, Kang},
  journal={Joule},
  volume={4},
  number={1},
  pages={69--100},
  year={2020},
  publisher={Elsevier}
}

@article{suo2017water,
  title={“Water-in-salt” electrolyte makes aqueous sodium-ion battery safe, green, and long-lasting},
  author={Suo, Liumin and Borodin, Oleg and Wang, Yuesheng and Rong, Xiaohui and Sun, Wei and Fan, Xiiulin and Xu, Shuyin and Schroeder, Marshall A and Cresce, Arthur V and Wang, Fei and others},
  journal={Advanced Energy Materials},
  volume={7},
  number={21},
  pages={1701189},
  year={2017},
  publisher={Wiley Online Library}
}

@article{Wang2018,
author = {Wang, Fei and Borodin, Oleg and Gao, Tao and Fan, Xiulin and Sun, Wei and Han, Fudong and Faraone, Antonio and Dura, Joseph A. and Xu, Kang and Wang, Chunsheng},
doi = {10.1038/s41563-018-0063-z},
issn = {14764660},
journal = {Nature Materials},
number = {6},
pages={543--549},
title = {{Highly reversible zinc metal anode for aqueous batteries}},
volume = {17},
year = {2018}
}

@article{smith2016electrostatic,
  title={The electrostatic screening length in concentrated electrolytes increases with concentration},
  author={Smith, Alexander M and Lee, Alpha A and Perkin, Susan},
  journal={J. Phys. Chem. Lett.},
  volume={7},
  number={12},
  pages={2157--2163},
  year={2016},
  publisher={ACS Publications}
}

@article{avni2020charge,
  title={Charge oscillations in ionic liquids: A microscopic cluster model},
  author={Avni, Yael and Adar, Ram M and Andelman, David},
  journal={Physical Review E},
  volume={101},
  number={1},
  pages={010601},
  year={2020},
  publisher={APS}
}

@article{mceldrew2020theory,
  title={Theory of Ion Aggregation and Gelation in Super-Concentrated Electrolytes},
  author={McEldrew, Michael and Goodwin, Zachary AH and Bi, Sheng and Bazant, Martin Z and Kornyshev, Alexei A},
  journal={J. Chem. Phys.},
  volume={152},
  pages={234506},
  year={2020}
}

@article{li2020new,
  title={New Concepts in Electrolytes},
  author={Li, Matthew and Wang, Chunsheng and Chen, Zhongwei and Xu, Kang and Lu, Jun},
  journal={Chem. Rev.},
  year={2020},
  publisher={ACS Publications}
}

@article{Cheng2022Sol,
  title={Emerging Era of Electrolyte Solvation
Structure and Interfacial Model in Batteries},
  author={Haoran Cheng and Qujiang Sun and Leilei Li and Yeguo Zou and Yuqi Wang and Tao Cai and Fei Zhao and Gang Liu and Zheng Ma and Wandi Wahyudi and Qian Li and Jun Ming},
  journal={ACS Energy Lett.},
  volume={7},
  pages={490--513},
  year={2022},
}

@article{Xie2023,
  title={Spatially resolved structural order in low-temperature liquid electrolyte},
  author={Yujun Xie and Jingyang Wang and Benjamin H. Savitzky and Zheng Chen and Yu Wang and Sophia Betzler and Karen Bustillo and Kristin Persson and Yi Cui and Lin-Wang Wang and Colin Ophus and Peter Ercius and Haimei Zheng},
  journal={Sci. Adv.},
  volume={9},
  pages={eadc9721},
  year={2023},
}

@article{Borodin2014SEI,
  title={Interfacial Structure and Dynamics of the Lithium Alkyl Dicarbonate SEI Components in Contact with the Lithium Battery Electrolyte},
  author={Oleg Borodin and Dmitry Bedrov},
  journal={J. Phys. Chem. C},
  volume={118},
  pages={18362--18371},
  year={2014},
}

@article{yu2020asymmetric,
  title={Asymmetric Composition of Ionic Aggregates and the Origin of High Correlated Transference Number in Water-in-Salt Electrolytes},
  author={Yu, Zhou and Curtiss, Larry A and Winans, Randall E and Zhang, Yang and Li, Tao and Cheng, Lei},
  journal={The Journal of Physical Chemistry Letters},
  volume={11},
  number={4},
  pages={1276--1281},
  year={2020},
  publisher={ACS Publications}
}

@article{molinari2020chelation,
  title={Chelation-Induced Reversal of Negative Cation Transference Number in Ionic Liquid Electrolytes},
  author={Molinari, Nicola and Kozinsky, Boris},
  journal={J. Phys. Chem. B},
  volume={124},
  number={13},
  pages={2676--2684},
  year={2020},
  publisher={ACS Publications}
}

@article{dou2018safe,
  title={Safe and high-rate supercapacitors based on an ``acetonitrile/water in salt” hybrid electrolyte},
  author={Dou, Qingyun and Lei, Shulai and Wang, Da-Wei and Zhang, Qingnuan and Xiao, Dewei and Guo, Hongwei and Wang, Aiping and Yang, Hui and Li, Yongle and Shi, Siqi and Yan, Xingbin},
  journal={Energy Environ. Sci.},
  volume={11},
  number={11},
  pages={3212--3219},
  year={2018},
  publisher={Royal Society of Chemistry}
}

@article{Zheng2017Uni,
  title={Research Progress towards Understanding the Unique
Interfaces between Concentrated Electrolytes and
Electrodes for Energy Storage Applications},
  author={Jianming Zheng and Joshua A. Lochala and Alexander Kwok and Zhiqun Daniel Deng and Jie Xiao},
  journal={Adv. Sci.},
  volume={4},
  pages={1700032},
  year={2017},
}

@article{xu2014electrolytes,
  title={Electrolytes and interphases in Li-ion batteries and beyond},
  author={Xu, Kang},
  journal={Chem. Rev.},
  volume={114},
  number={23},
  pages={11503--11618},
  year={2014},
  publisher={ACS Publications}
}

@article{adar2019screening,
  title={Screening length for finite-size ions in concentrated electrolytes},
  author={Adar, Ram M and Safran, Samuel A and Diamant, Haim and Andelman, David},
  journal={Physical Review E},
  volume={100},
  number={4},
  pages={042615},
  year={2019},
  publisher={APS}
}

@article{krucker2021underscreening,
  title={Underscreening and hidden ion structures in large scale simulations of concentrated electrolytes},
  author={Krucker-Velasquez, Emily and Swan, James W},
  journal={The Journal of Chemical Physics},
  volume={155},
  number={13},
  pages={134903},
  year={2021},
  publisher={AIP Publishing}
}

@article{Zhang2024,
      title={Long-Range Surface Forces in Salt-in-Ionic Liquids},
      author={Zhang, Xuhui and Goodwin, Zachary AH and Hoane, Alexis G and Deptula, Alex and Markiewitz, Daniel M and Molinari, Nicola and Zheng, Qianlu and Li, Hua and McEldrew, Michael and Kozinsky, Boris and others},
      journal={ACS nano},
      volume={18},
      number={50},
      pages={34007--34022},
      year={2024},
      publisher={ACS Publications}
}

@article{markiewitz2024,
  title={Electric field induced associations in the double layer of salt-in-ionic-liquid electrolytes},
  author={Markiewitz, Daniel M and Goodwin, Zachary AH and McEldrew, Michael and de Souza, J Pedro and Zhang, Xuhui and Espinosa-Marzal, Rosa M and Bazant, Martin Z},
  journal={Faraday Discussions},
  pages={365--384},
  year={2024},
  publisher={Royal Society of Chemistry}
}

@article{efaw2023localized,
  title={Localized high-concentration electrolytes get more localized through micelle-like structures},
  author={Efaw, Corey M and Wu, Qisheng and Gao, Ningshengjie and Zhang, Yugang and Zhu, Haoyu and Gering, Kevin and Hurley, Michael F and Xiong, Hui and Hu, Enyuan and Cao, Xia and others},
  journal={Nature Materials},
  volume={22},
  number={12},
  pages={1531--1539},
  year={2023},
  publisher={Nature Publishing Group UK London}
}

@article{groves2021surface,
  title={Surface forces and structure in a water-in-salt electrolyte},
  author={Groves, Timothy S and Perez-Martinez, Carla S and Lhermerout, Romain and Perkin, Susan},
  journal={The Journal of Physical Chemistry Letters},
  volume={12},
  number={6},
  pages={1702--1707},
  year={2021},
  publisher={ACS Publications}
}

@article{hartel2023anomalous,
  title={Anomalous underscreening in the restricted primitive model},
  author={H{\"a}rtel, Andreas and B{\"u}ltmann, Moritz and Coupette, Fabian},
  journal={Physical Review Letters},
  volume={130},
  number={10},
  pages={108202},
  year={2023},
  publisher={APS}
}

@article{perez2017underscreening,
  title={Underscreening in concentrated electrolytes},
  author={A Lee and Perez-Martinez, Carla S and Smith, Alexander M and Perkin, Susan and others},
  journal={Faraday discussions},
  volume={199},
  pages={239--259},
  year={2017},
  publisher={Royal Society of Chemistry}
}

@misc{Goodenough2013ThePerspective,
    title = {{The Li-ion rechargeable battery: A perspective}},
    year = {2013},
    booktitle = {Journal of the American Chemical Society},
    author = {Goodenough, John B. and Park, Kyu Sung},
    number = {4},
    volume = {135},
    doi = {10.1021/ja3091438},
    issn = {00027863}
}

@article{phelan2024role,
  title={Role of salt concentration in stabilizing charged Ni-rich cathode interfaces in Li-ion batteries},
  author={Phelan, Conor ME and Bj\"orklund, Erik and Singh, Jasper and Fraser, Michael and Didwal, Pravin N and Rees, Gregory J and Ruff, Zachary and Ferrer, Pilar and Grinter, David C and Grey, Clare P and others},
  journal={Chemistry of Materials},
  volume={36},
  number={7},
  pages={3334--3344},
  year={2024},
  publisher={ACS Publications}
}

\end{document}